\documentclass[a4paper,11pt]{article}
\pdfoutput=1
\usepackage[utf8]{inputenc}
\usepackage{jcappub}
\usepackage[T1]{fontenc}
\usepackage{amsmath}

\newcommand{\Tend}{T_\text{end}}
\newcommand{\Teq}{T_\text{eq}}
\newcommand{\Tfo}{T_\text{fo}}
\newcommand{\Tc}{T_\text{c}}
\newcommand{\aend}{a_\text{end}}
\newcommand{\aeq}{a_\text{eq}}
\newcommand{\ac}{a_\text{c}}
\newcommand{\afo}{a_\text{fo}}
\newcommand{\xfo}{x_\text{fo}}
\newcommand{\xeq}{x_\text{eq}}
\newcommand{\xc}{x_\text{c}}
\newcommand{\xend}{x_\text{end}}
\newcommand{\gs}{g_\star}
\newcommand{\gss}{g_{\star S}}
\newcommand{\sv}{\langle\sigma v\rangle}
\newcommand{\rp}{\rho_\phi}
\newcommand{\rR}{\rho_R}
\newcommand{\TBBN}{T_\text{BBN}}

\title{Reconstructing Non-standard Cosmologies with Dark Matter}

\author[a]{Paola Arias,}
\author[b]{Nicolás Bernal,}
\author[b,\,c]{\\Alan Herrera,}
\author[a]{Carlos Maldonado}

\affiliation[a]{Departmento de Física, Universidad de Santiago de Chile,\\
Casilla 307, Santiago, Chile}
\affiliation[b]{Centro de Investigaciones, Universidad Antonio Nariño,\\
Carrera 3 Este \# 47A-15, Bogotá, Colombia}
\affiliation[c]{
Department of Physics and Astronomy, University of Rochester,\\ Rochester, NY 14627, USA.}

\emailAdd{paola.arias.r@usach.cl}
\emailAdd{nicolas.bernal@uan.edu.co}
\emailAdd{aherrer9@ur.rochester.edu}
\emailAdd{carlos.maldonados@usach.cl}

\abstract{Once dark matter has been discovered and its particle physics properties have been determined, a crucial question rises concerning how it was produced in the early Universe.
If its thermally averaged annihilation cross section is in the ballpark of few$\times 10^{-26}$~cm$^3$/s, the WIMP mechanism in the standard cosmological scenario (i.e. radiation dominated Universe) will be highly favored.
If this is not the case one can either consider an alternative production mechanism, or a non-standard cosmology.
Here we study the dark matter production in scenarios with a non-standard expansion history.
Additionally, we reconstruct the possible non-standard cosmologies that could make the WIMP mechanism viable.
}

\begin{document}

\begin{flushright}
PI/UAN-2019-649FT
\end{flushright}

\maketitle

%%%%%%%%%%%%%%%%%%%%%%%%%%%%%%%%%%%%%%%%%%%%%%%%%%%%%%%%%%%%%%%%%%%%%%%%%%%%%%%
%%%%%%%%%%%%%%%%%%%%%%%%%%%%%%%%%%%%%%%%%%%%%%%%%%%%%%%%%%%%%%%%%%%%%%%%%%%%%%%
%%%%%%%%%%%%%%%%%%%%%%%%%%%%%%%%%%%%%%%%%%%%%%%%%%%%%%%%%%%%%%%%%%%%%%%%%%%%%%%
\section{Introduction}
There is compelling evidence for the existence of Dark Matter (DM), an unknown, non-baryonic matter component whose abundance in the Universe exceeds the amount of ordinary matter roughly by a factor of five~\cite{Aghanim:2018eyx}.
In the previous decades a class of scenarios where dark and visible matter were once in thermal equilibrium with each other has received by far the biggest attention, both theoretically and experimentally.
Most prominent in this class are extensions of the Standard Model of particle physics (SM) that feature Weakly Interacting Massive Particles (WIMPs) as DM~\cite{Bertone:2004pz, Arcadi:2017kky, Lin:2019uvt, Hooper:2018kfv}.

Despite the fact that WIMP DM has been searched for decades, the studies have yielded no overwhelming evidence for what DM actually is.
A crucial challenge to the WIMP DM paradigm is the lack of a confirmed experimental detection signal.
The worldwide program for detecting WIMP DM using a multi-channel and multi-messenger approach has followed three main strategies: direct detection, indirect detection, and production at colliders.

However, the observed DM abundance may have been generated also out of equilibrium by a mechanism like the so-called freeze-in~\cite{McDonald:2001vt, Choi:2005vq, Kusenko:2006rh, Petraki:2007gq, Hall:2009bx, Chu:2011be, Belanger:2018mqt} (for a recent review see ref.~\cite{Bernal:2017kxu}).
Another simple way to evade the experimental constraints on DM is to consider non-standard cosmological histories, for example scenarios where the Universe was effectively matter-dominated at an early stage, due to a slow reheating period after inflation or to a massive metastable particle.
There are no reasons to assume that the Universe was radiation-dominated prior to Big Bang Nucleosynthesis (BBN).\footnote{For studies on baryogenesis with a low reheating temperature or during an early matter-dominated phase, see refs.~\cite{Davidson:2000dw, Giudice:2000ex, Allahverdi:2010im, Beniwal:2017eik, Allahverdi:2017edd} and~\cite{Bernal:2017zvx}, respectively.
Additionally, primordial gravitational wave production in scenarios with an early matter era have recently received particular attention~\cite{Assadullahi:2009nf, Durrer:2011bi, Alabidi:2013wtp, DEramo:2019tit, Bernal:2019lpc, Figueroa:2019paj}.}

Examples of non-standard cosmologies are abundant in the literature.
For instance, in typical string theory models there are many scalar moduli fields, the mass of which is typically set by the gravitino mass.
If this is fairly low, motivated for example by the success of gauge unification in supersymmetric extensions of the SM, the moduli naturally dominate the energy density of the Universe at early times leading to an extended period of matter domination.
The moduli eventually decay through Planck suppressed operators, and a radiation dominated Universe re-emerges before BBN.
Additionally, in the kination scenario~\cite{Barrow:1982ei, Ford:1986sy} $\phi$ is a `fast-rolling' field whose kinetic energy governs the expansion rate of the post-inflation Universe, with an equation of state $\omega=1$.
Due to the scaling of the energy density in radiation with the scale factor $\rho_R\propto a^{-4}$, which is slower than the scaling of the energy density in the $\phi$ field $\rho_\phi\propto a^{-6}$, the contribution from the radiation energy density in determining the expansion rate eventually becomes more important than that from the $\phi$ field.
When the $\phi$ field redshifts away, the standard radiation dominated cosmology takes place.
In general, production of DM in scenarios with a non-standard expansion phase has recently gained increasing interest, see e.g. refs.~\cite{Kane:2015jia, Co:2015pka, Davoudiasl:2015vba, Randall:2015xza, Berlin:2016vnh, Tenkanen:2016jic, Dror:2016rxc, Berlin:2016gtr, DEramo:2017gpl, Hamdan:2017psw, Visinelli:2017qga, Dror:2017gjq, Drees:2017iod, DEramo:2017ecx, Maity:2018dgy, Bernal:2018ins, Hardy:2018bph, Maity:2018exj, Hambye:2018qjv, Bernal:2018kcw, Arbey:2018uho, Drees:2018dsj, Betancur:2018xtj, Maldonado:2019qmp, Poulin:2019omz, Tenkanen:2019cik}.
For earlier works, see also refs.~\cite{Barrow:1982ei, Kamionkowski:1990ni, McDonald:1989jd, Salati:2002md, Comelli:2003cv, Rosati:2003yw, Pallis:2004yy, Gelmini:2006pw, Gelmini:2006pq, Arbey:2008kv, Cohen:2008nb, Arbey:2009gt}.
Additionally, a non-standard period might have lasted for a considerable amount of time, namely since the end of inflation down to the moment when BBN started~\cite{Chung:1998rq, Giudice:2000ex, Kolb:2003ke, Garcia:2017tuj, Ellis:2015jpg}.
In these modified cosmologies, various properties of the WIMPs like their free-streaming velocity and the temperature at which the kinetic decoupling occurs have been investigated~\cite{Gelmini:2008sh, Visinelli:2015eka, Waldstein:2016blt, Waldstein:2017wps}.

If DM is a WIMP that is a thermal relic of the early Universe, then its total thermally averaged self-annihilation cross section $\sv$ is revealed by its present-day mass density.
In standard cosmological scenarios, this result for a generic WIMP is usually stated as $\sv_0=\text{few}\times 10^{-26}$~cm$^3$/s$~=\text{few}\times 10^{-9}$~GeV$^{-2}$, with a small logarithmic dependence of WIMP mass~\cite{Steigman:2012nb}.
If $\sv\gg\sv_0$, DM is kept in chemical equilibrium with the thermal bath for longer, giving rise to a DM underabundance that can be understood for example in the context of multicomponent DM.
On the contrary, if $\sv\ll\sv_0$, DM decouples earlier and generates an overabundance that overcloses the Universe.
In non-standard cosmologies, however, the generic value for $\sv_0$ does not hold anymore, strongly depending on the details of the cosmology.

Once DM is discovered and its particle physics properties have been reconstructed (i.e. mass and couplings with the SM), a major question rises concerning the DM production mechanism.\footnote{It is necessary to make use of the complementarity between different experiments and different detection techniques~\cite{Mena:2007ty, Drees:2008bv, Bernal:2008zk, Bernal:2008cu, Bergstrom:2010gh, Pato:2010zk, Arisaka:2011eu, Cerdeno:2013gqa, Arina:2013jya, Peter:2013aha, Kavanagh:2014rya, Roszkowski:2016bhs, Queiroz:2016sxf, Roszkowski:2017dou, Kavanagh:2017hcl, Bertone:2017adx, Queiroz:2018utk} in order to ameliorate
determination of the particle physics parameters and disentangle possible degeneracies.
Furthermore, one has to take into account astrophysical uncertainties~\cite{Green:2002ht, Zemp:2008gw, McCabe:2010zh, Bernal:2010ip,Pato:2010yq, Bernal:2011pz, Fairbairn:2012zs, Bernal:2014mmt, Bernal:2015oyn, Bernal:2016guq, Benito:2016kyp, Green:2017odb, Ibarra:2018yxq, Benito:2019ngh, Karukes:2019jxv, Wu:2019nhd} when interpreting the results of the DM searches.}
If the inferred value of $\sv$ is in the ballpark of $\sv_0$, the simpler freeze-out mechanism with a standard cosmology will be strongly favored.
However, if that turns out not to be the case, one can either look for different DM production mechanisms or for alternative cosmological scenarios.
The latter option will be pursued in this study.

In this paper we consider production of WIMP DM in scenarios where for some period at early times (for temperatures around the DM mass) the expansion of the Universe was governed by a component $\phi$ with an effective equation of state $\omega=p_\phi/\rp$, where $p_\phi$ is the pressure and $\rp$ the energy density of $\phi$.
Using a particle physics model independent approach, for a given DM mass $m$ and a thermally averaged DM annihilation cross section $\sv$, we study the capabilities  for reconstructing the parameters characterizing the non-standard cosmology.
The paper is organized as follows:
In section~\ref{sec:NonStandard} we introduce the cosmological setup.
In section~\ref{sec:Reconstruction} we present the reconstruction capabilities of the cosmological parameters.
Finally, we conclude in section~\ref{sec:Conclusion}.

%%%%%%%%%%%%%%%%%%%%%%%%%%%%%%%%%%%%%%%%%%%%%%%%%%%%%%%%%%%%%%%%%%%%%%%%%%%%%%%
%%%%%%%%%%%%%%%%%%%%%%%%%%%%%%%%%%%%%%%%%%%%%%%%%%%%%%%%%%%%%%%%%%%%%%%%%%%%%%%
%%%%%%%%%%%%%%%%%%%%%%%%%%%%%%%%%%%%%%%%%%%%%%%%%%%%%%%%%%%%%%%%%%%%%%%%%%%%%%%
\section{Non-Standard Cosmologies}\label{sec:NonStandard}
We assume that for some period of the early Universe, the total energy density was dominated by a component $\rp$ with an equation of state parameter $\omega$, where $\omega\equiv p_\phi/\rp$, with $p_\phi$ the pressure of the dominant component.
Additionally, this component decays with a total rate $\Gamma_\phi$.

In the early Universe the evolution of the $\phi$ energy density $\rp$, the SM entropy density $s$, as well as the DM number density $n$ are governed by the system of coupled Boltzmann equations~\cite{Giudice:2000ex,Drees:2017iod}
\begin{align}
\frac{d\rp}{dt}+3(1+\omega)\,H\,\rp&=-\Gamma_\phi\,\rp\,,\label{eq:cosmo2} \\
\frac{ds}{dt}+3\,H\,s&=+\frac{\Gamma_\phi\,\rp}{T}\left(1-b\frac{E}{m_\phi}\right)+2\frac{E}{T}\sv\left(n^2-n_\text{eq}^2\right),\label{eq:cosmo3}\\
\frac{dn}{dt}+3\,H\,n&=+\frac{b}{m_{\phi}}\Gamma_\phi\,\rho_\phi-\sv\left(n^2-n_\text{eq}^2\right),\label{eq:cosmo1}
\end{align}
where $\sv$ is the total DM annihilation cross-section into SM particles and $E^2\simeq m^2+3\,T^2$ is the averaged energy per DM particle.
In general $\phi$ decays into both SM radiation and DM particles~\cite{Kane:2015qea}, with a proportion controlled by the parameter $b$.
In fact, $b$ is twice the branching ratio of $\phi$ decaying into a couple of DM particles\footnote{We assume here that the main decay channel of $\phi$ into DM particles is into two of them.} and $m_\phi$ corresponds to the mass of the state $\phi$.
Additionally, $1-b\,E/m_\phi$ is the fraction of $\rp$ that goes into radiation.
The second term in the RHS of eq.~\eqref{eq:cosmo3} corresponding to the entropy injection due to DM annihilations is subdominant and thus is ignored.

Additionally, the two terms in the RHS of eq.~\eqref{eq:cosmo1} represent the non-thermal production via the decay of $\phi$, and the usual thermal WIMP production, respectively.
However, here we focus in the case where DM is thermally produced, which implies that the branching ratio of $\phi$ into DM particles is subdominant, and therefore we disregard it, i.e. $b=0$.%
\footnote{Let us note that the decay into DM particles can be disregarded as long as $b<10^{-4}\,m/(100$~GeV)~\cite{Drees:2017iod}.}

Under the assumption that the SM plasma maintains internal equilibrium at all times in the early Universe, its temperature dependence can be obtained from its energy density
\begin{equation}\label{eq:rhoSM}
    \rR(T)=\frac{\pi^2}{30}\,\gs(T)\,T^4.
\end{equation}
Equation~\eqref{eq:cosmo3} plays an important role in tracking the evolution of the photon's temperature $T$, via the SM entropy density $s$,
\begin{equation}
    s(T)=\frac{\rR+p_R}{T}=\frac{2\pi^2}{45}\,\gss(T)\,T^3,
\end{equation}
where $\gs(T)$ and $\gss(T)$ correspond to the effective number of relativistic degrees of freedom for SM energy and entropy densities, respectively~\cite{Drees:2015exa}. 
The evolution of the SM temperature follows from Eq.~\eqref{eq:cosmo3}:
\begin{equation}\label{eq:cosmo3b}
    \frac{dT}{da}=\left(1+\frac{T}{3\,\gss}\frac{d\gss}{dT}\right)^{-1}\left[-\frac{T}{a}+\frac{\Gamma_\phi\,\rp}{3\,H\,s\,a}\left(1-\frac{E\,b}{m_\phi}\right)+\frac23\frac{E\,\sv}{H\,s\,a}\left(n^2-n_\text{eq}^2\right)\right].
\end{equation}
The Hubble expansion rate $H$ is defined by
\begin{equation}
    H^2=\frac{\rp+\rR+\rho_\chi}{3\,M_P^2},
\end{equation}
where $M_P$ is the reduced Planck mass.

For having a successful BBN, the temperature at the end of the $\rp$ dominated phase has to be $\Tend\gtrsim 4$~MeV~\cite{Kawasaki:2000en, Hannestad:2004px, Ichikawa:2005vw, DeBernardis:2008zz}, where $\Tend$ is given by the total decay width $\Gamma_\phi$ as
\begin{equation}\label{eq:Tend}
    \Tend^4\equiv\frac{90}{\pi^2\,\gs(\Tend)}\,M_P^2\,\Gamma_\phi^2\,.
\end{equation}
Let us note that for $\omega>1/3$, $\rp$ gets diluted faster than radiation, and if $\rp\ll\rR$ at $T=\TBBN$, $\Gamma_\phi$ could be effectively taken to be zero.

\begin{figure}[t]
\begin{center}
\includegraphics[height=0.33\textwidth]{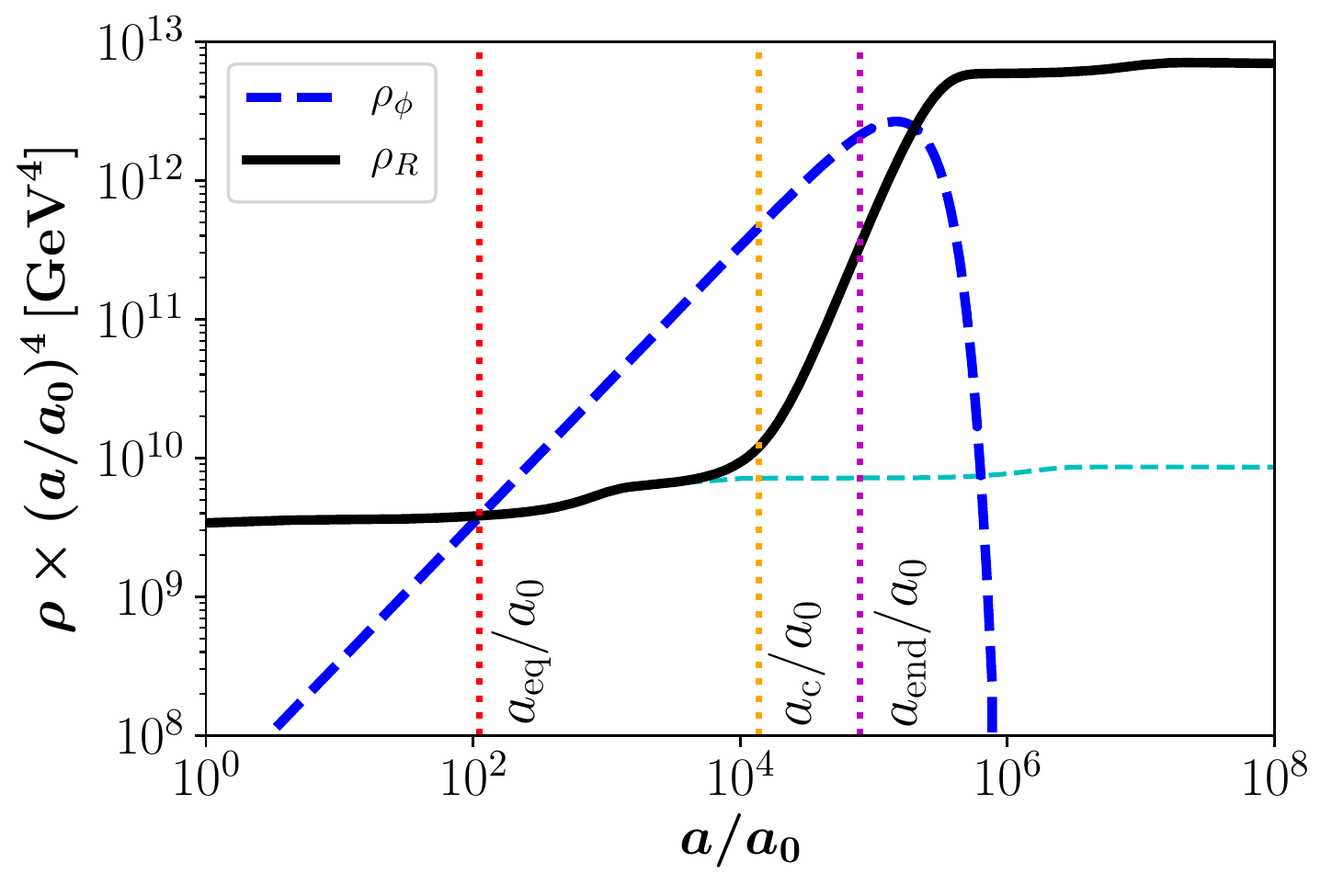}
\includegraphics[height=0.33\textwidth]{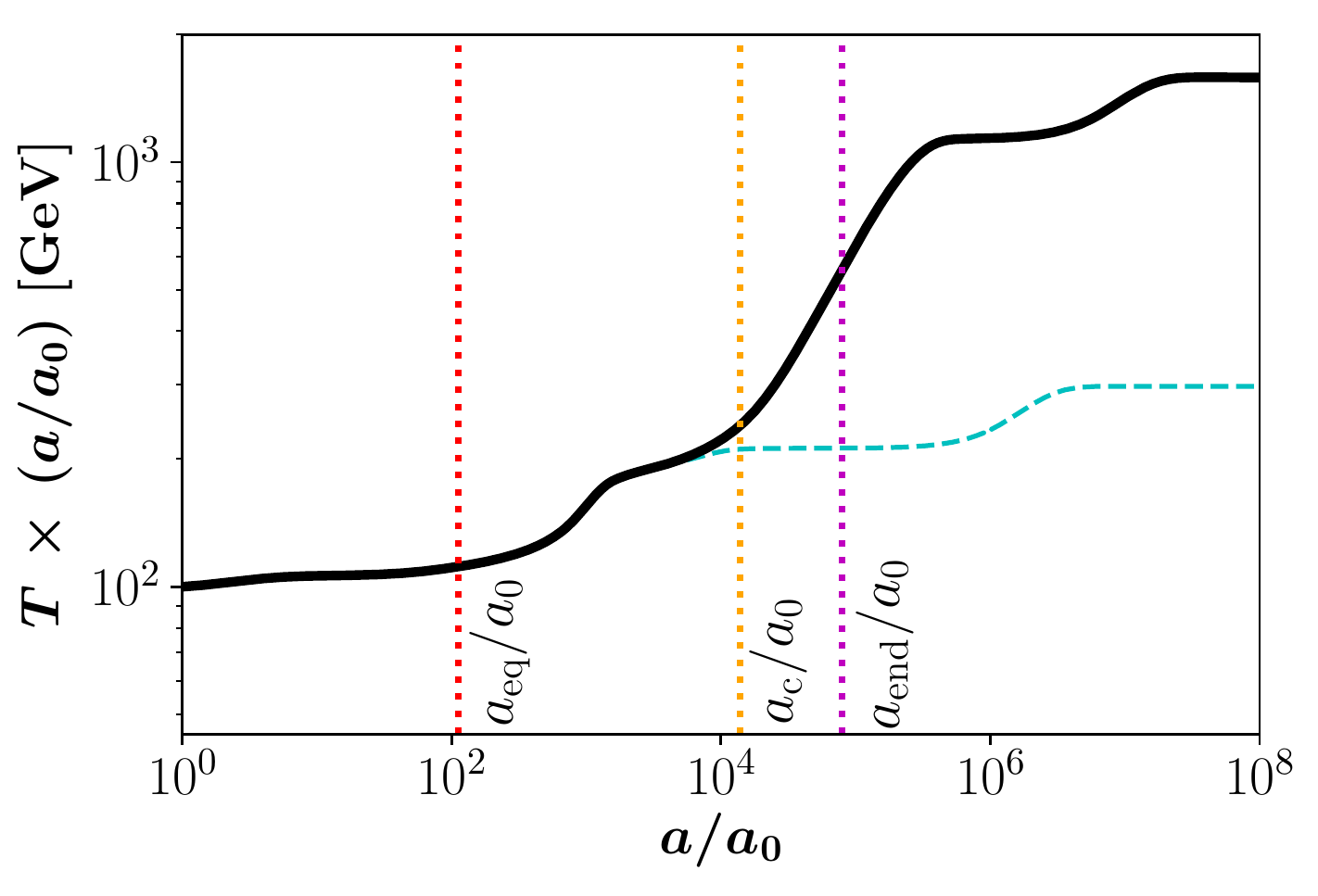}
\caption{Evolution of the energy densities for radiation and the $\phi$ field (left panel), and evolution of the photon temperature $T$ (right panel) as a function of the scale factor $a$, for $\omega=0$, $\Tend=7\times 10^{-3}$~GeV and $\left.\frac{\rp}{\rR}\right|_{T=100~\text{GeV}}=10^{-2}$.
The cyan dashed lines represent the standard cosmological scenario without the $\phi$ field.
Here we assume that $a_0= a(T=100~\text{GeV})$.
The scale factors $a=\aeq$, $\ac$ and $\aend$ are overlaid.
}
\label{fig:rho}
\end{center}
\end{figure}
As an example, fig.~\ref{fig:rho} shows the solution of the Boltzmann equations~\eqref{eq:cosmo2} and~\eqref{eq:cosmo3}, for $\omega=0$, $\Tend=7\times 10^{-3}$~GeV and $\left.\frac{\rp}{\rR}\right|_{T=100~\text{GeV}}=10^{-2}$.%
\footnote{Let us emphasize that all the figures in this work were produced using the full numerical expressions.}
The left panel depicts the evolution of the radiation and $\phi$ energy densities as a function of the scale factor $a$, taking $a_0= a(T=100~\text{GeV})$.
$\aeq$ corresponds to the scale factor at which $\rp$ starts to dominate over $\rR$,
$\ac$ to the scale factor where effectively $\rp$ starts to dominate the evolution of $\rR$, and $\aend$ is a proxy of the scale factor where $\phi$ decays completely.
Additionally $\Teq\equiv T(a=\aeq)$, $\Tc\equiv T(a=\ac)$ and $\Tend\equiv T(a=\aend)$.
$\Tend$ is properly defined in eq.~\eqref{eq:Tend}.
The photon temperature can be extracted from the radiation energy density using eq.~\eqref{eq:rhoSM} and it is depicted as a function of the scale factor in the right panel of fig.~\ref{fig:rho}.
The bumps at $a/a_0\sim10^3$ and $\sim10^7$, corresponding to temperatures $T\sim 10^{-1}$~GeV and $\sim10^{-3}$~GeV, are due to the QCD phase transition and the annihilation of electron-positron pairs, respectively.
For completeness, we also show in the figure with cyan dashed lines the evolution of the SM energy density and the temperature in the case without the $\phi$ field. 
In the case with constant relativistic degrees of freedom one has that $\rp(a)\propto a^{-3(1+\omega)}$ until it decays, and
\begin{equation}
    \rR(a)\propto
    \begin{cases}
    a^{-4}\qquad\qquad \text{ for }\hspace{1.25cm}a\ll\ac,\\
    a^{-\frac32(1+\omega)}\hspace{0.8cm} \text{for }\ac\hspace{0.35cm}\ll a\ll\aend,\\
    a^{-4}\qquad\qquad \text{ for }\aend\ll a,
    \end{cases}
\end{equation}
which implies that
\begin{equation}\label{eq:temperature}
    T(a)\propto
    \begin{cases}
    a^{-1}\qquad\qquad \text{ for }\hspace{1.25cm}a\ll\ac,\\
    a^{-\frac38(1+\omega)}\hspace{0.8cm} \text{for }\ac\hspace{0.35cm}\ll a\ll\aend,\\
    a^{-1}\qquad\qquad \text{ for }\aend\ll a.
    \end{cases}
\end{equation}

%%%%%%%%%%%%%%%%%%%%%%%%%%%%%%%%%%%%%%%%%%%%%%%%%%%%%%%%%%%%%%%%%%%%%%%%%%%%%%%
%%%%%%%%%%%%%%%%%%%%%%%%%%%%%%%%%%%%%%%%%%%%%%%%%%%%%%%%%%%%%%%%%%%%%%%%%%%%%%%
%%%%%%%%%%%%%%%%%%%%%%%%%%%%%%%%%%%%%%%%%%%%%%%%%%%%%%%%%%%%%%%%%%%%%%%%%%%%%%%
\section{Reconstructing Cosmological Parameters}\label{sec:Reconstruction}

In order to have a successful WIMP mechanism, a thermally averaged annihilation cross section $\sv\sim$ few$\times 10^{-26}$~cm$^3/$s is typically needed~\cite{Steigman:2012nb}.
If a DM measurement points towards a significantly different value, the simplest WIMP scenario could still be the responsible for the DM genesis, but with a non-standard cosmological evolution.

Here, we assume that both the DM mass $m$ and its thermally averaged annihilation cross section $\sv$ are known after a discovery, and we try to reconstruct the non-standard cosmological parameters that make the DM genesis compatible with the WIMP paradigm.
In this study we consider scenarios where for some period at early times the expansion of the Universe was governed by a fluid component with an effective equation of state $\omega$.
Particular cases correspond to $\omega=-1$ (quintessence), 0 (dust), 1/3 (radiation) and 1 (kination).
However, we will mainly focus on a phase of matter domination  assuming $\omega=0$.

The non-standard cosmologies considered here can be fully parametrized with three free parameters:
\begin{equation}
\Tend,\quad\kappa\equiv\left.\frac{\rp}{\rR}\right|_{T=m}\quad\text{and}\quad\omega\,.
\end{equation}

\begin{figure}[t]
\begin{center}
\includegraphics[height=0.4\textwidth]{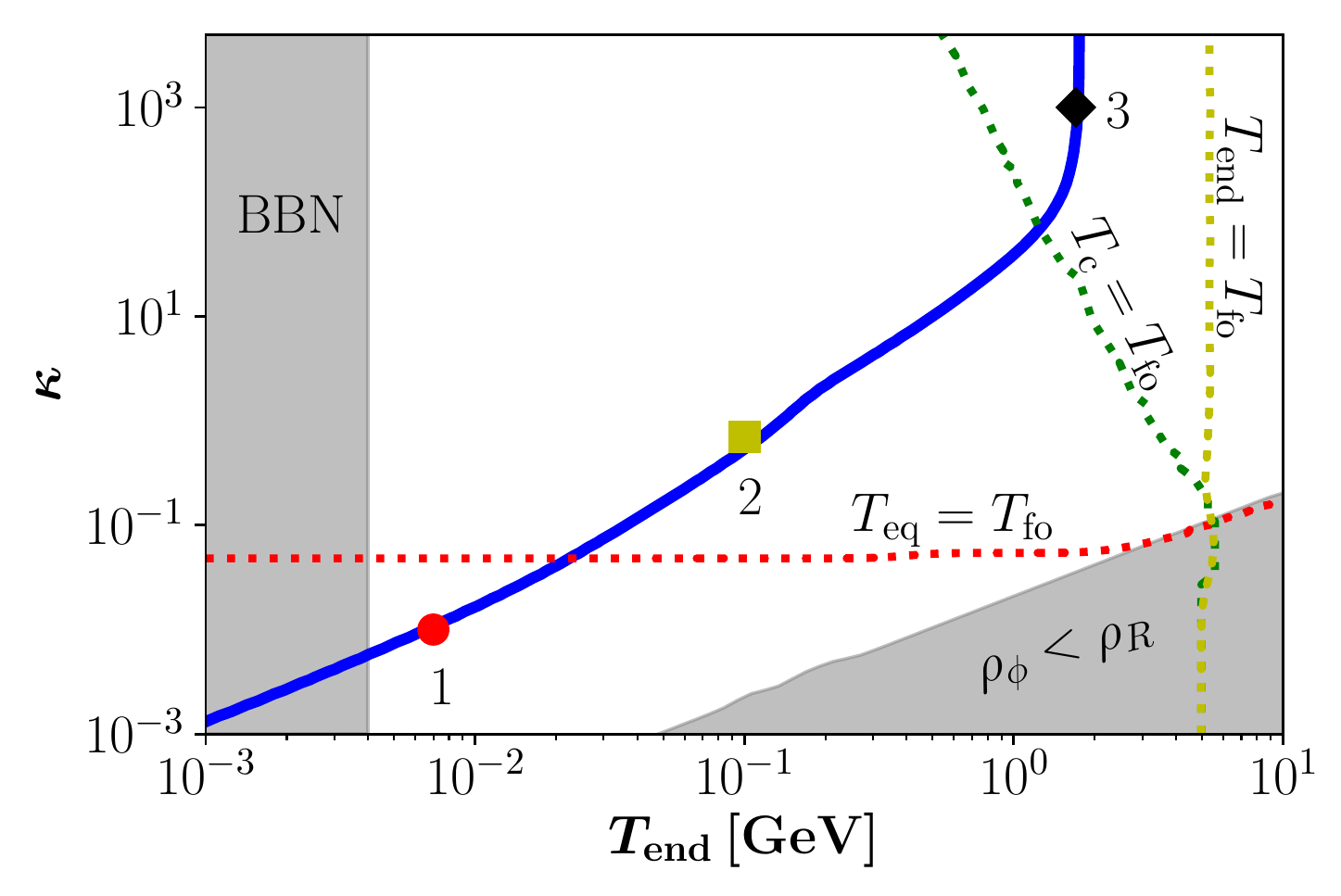}
\caption{Parameter space generating the observed DM abundance via the WIMP mechanism with non-standard cosmologies, assuming $\omega=0$.
For the particle physics benchmark we have taken $m=100$~GeV and $\sv=10^{-11}$~GeV$^{-2}$.
The gray bands correspond to $\Tend<\TBBN$ and $\rp<\rR$.
Benchmarks 1, 2 and 3 are the parameters used in figs.~\ref{fig:yield} and~\ref{fig:pre_sv-m}.
The lines corresponding to $\Tfo=\Teq$, $\Tfo=\Tc$ and $\Tfo=\Tend$ are overlaid.
}
\label{fig:reconstruction}
\end{center}
\end{figure}
Figure~\ref{fig:reconstruction} depicts in blue the parameter space compatible with the observed DM abundance via the WIMP mechanism with non-standard cosmologies, in the plane $[\Tend,\,\kappa]$, assuming $\omega=0$.
For the particle physics benchmark we have chosen $m=100$~GeV and $\sv=10^{-11}$~GeV$^{-2}$.
The left part of the plot colored in gray and corresponding to $\Tend<4$~MeV is in tension with BBN.
Additionally, in the lower right corner $\rp$ is always subdominant with respect to radiation, and hence corresponds to the usual case, radiation dominated.%
\footnote{That region can be understood in the sudden decay approximation, where the equality $\rR(\aend)=\rp(\aend)$ takes place, implying that $\kappa=\frac{\gs(\Tend)}{\gs(m)}\left[\frac{\Tend}{m}\right]^{1-3\omega}$.}
The figure also shows the lines corresponding to $\Tfo=\Teq$, $\Tfo=\Tc$ and $\Tfo=\Tend$.
These lines differentiate four phenomenologically distinct regimes characterized by the temperature $\Tfo$ when the DM freeze-out happens, with respect to $\Teq$, $\Tc$ and $\Tend$.
These cases are described in detail in the next subsections, where analytic estimations of the different regimes are performed.%
\footnote{For the analytical estimations the variation of the number of relativistic degrees of freedom $\gs$ and $\gss$ is ignored.
Additionally, we will take $\gss=\gs$, which is a good approximation for $T>1$~MeV.}

%%%%%%%%%%%%%%%%%%%%%%%%%%%%%%%%%%%%%%%%%%%%%%%%%%%%%%%%%%%%%%%%%%%%%%%%%%%%%%%
\subsection{Classification}
%%%%%%%%%%%%%%%%%%%%%%%%%%%%%%%%%%%%%%%%%%%%%%%%%%%%%%%%%%%%%%%%%%%%%%%%%%%%%%%
\subsubsection[Case 1: $\Teq\ll\Tfo$]{Case 1: $\boldsymbol{\Teq\ll\Tfo}$}\label{sec:1}
\begin{figure}[ht!]
\begin{center}
\includegraphics[height=0.32\textwidth]{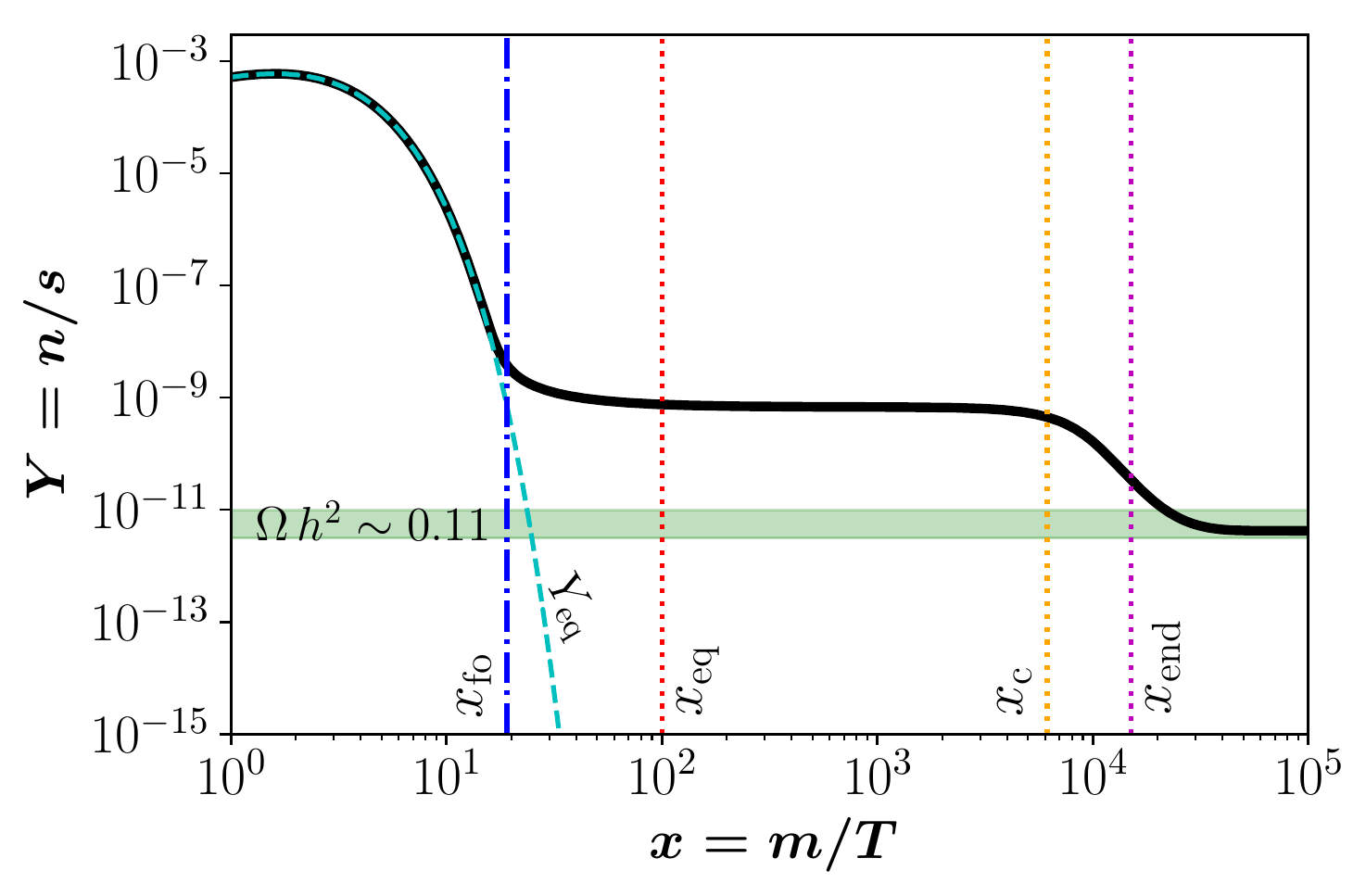}
\includegraphics[height=0.32\textwidth]{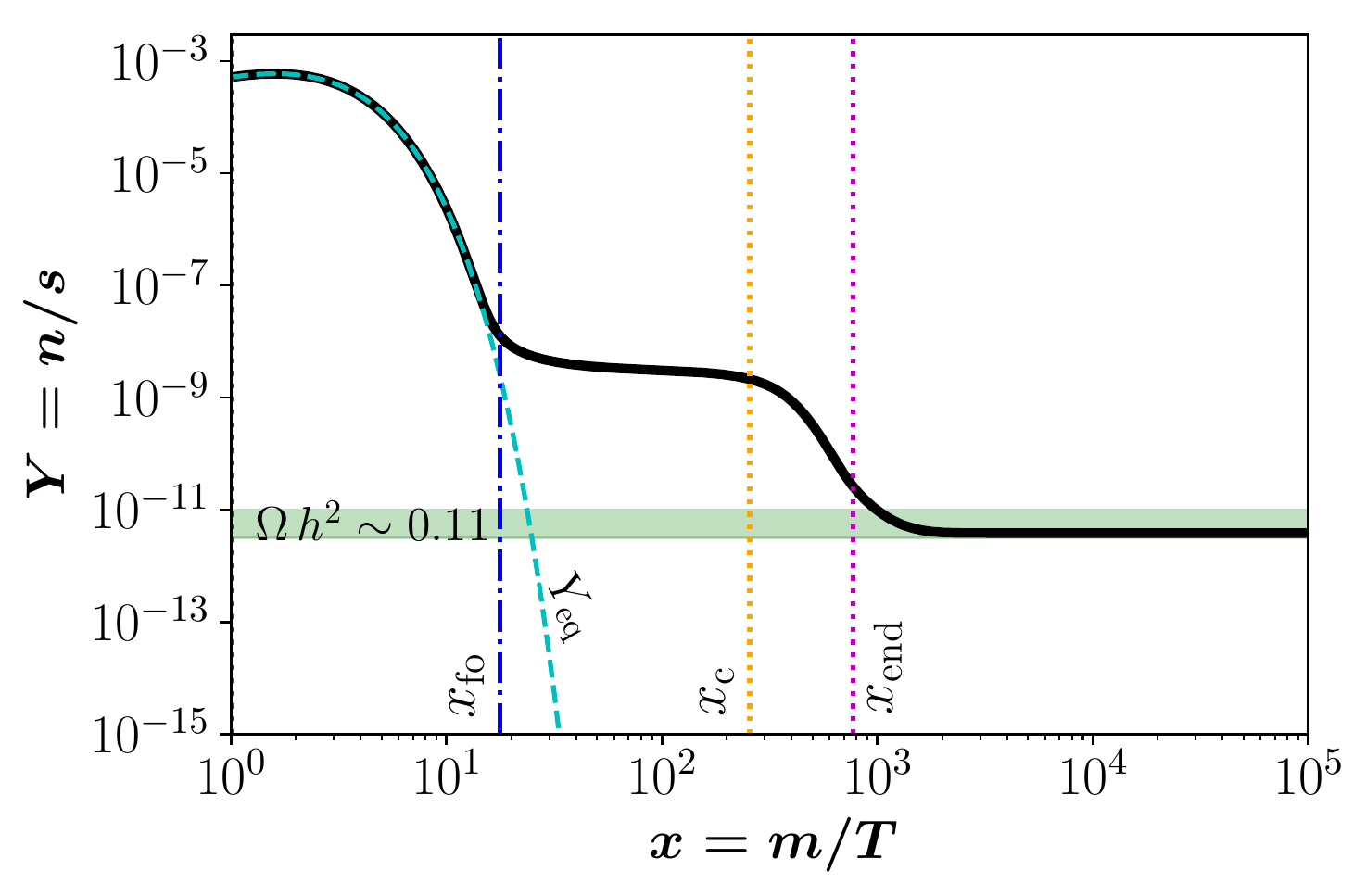}
\includegraphics[height=0.32\textwidth]{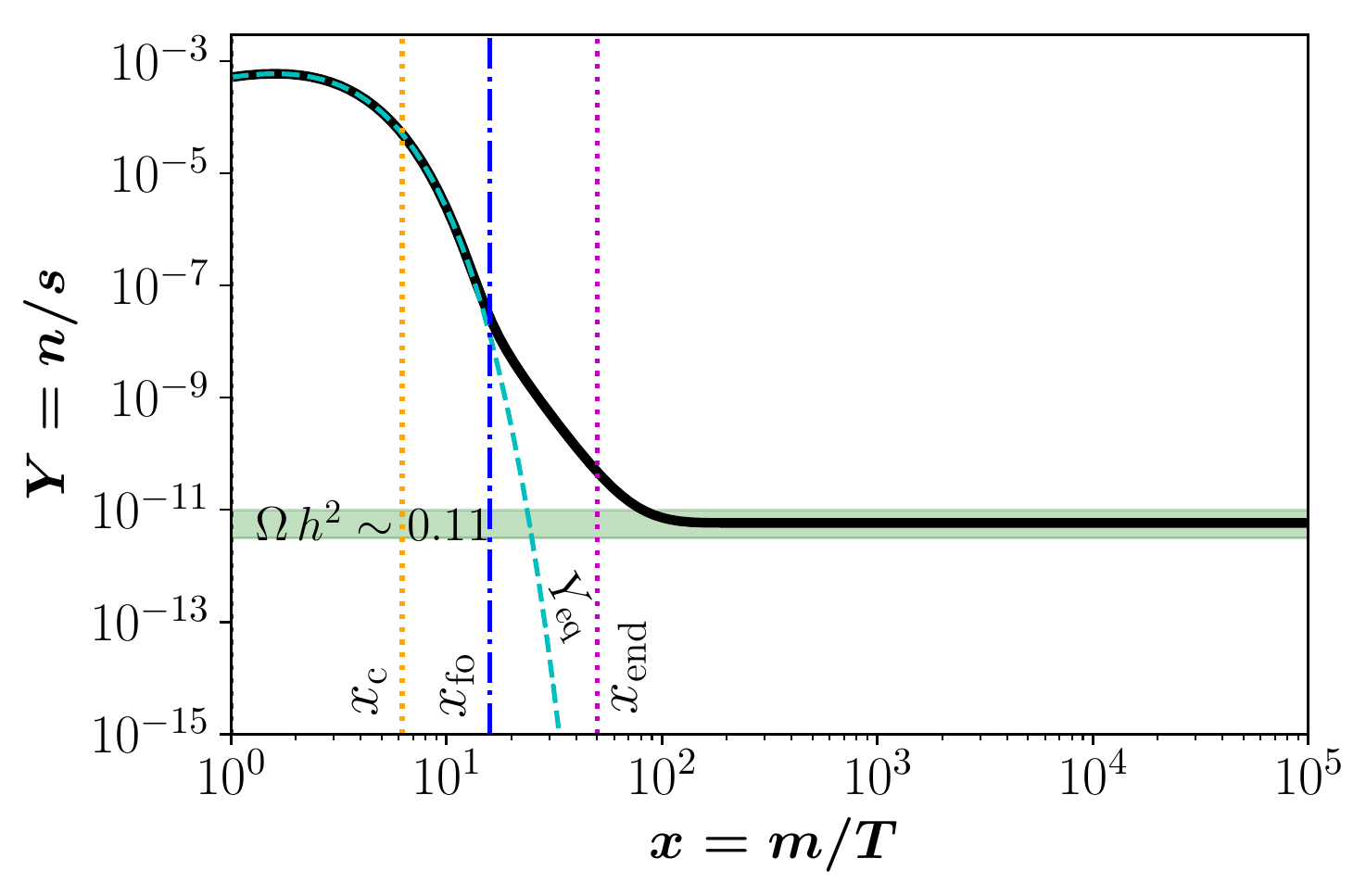}
\caption{Evolution of the DM yield (thick black lines) as a function of the inverse of the temperature using $m=100$~GeV and $\sv=10^{-11}$~GeV$^{-2}$, for the benchmark points shown if fig.~\ref{fig:reconstruction}.
Upper left panel: $\Tend=7\times 10^{-3}$~GeV and $\kappa=10^{-2}$. Upper right panel: $\Tend=10^{-1}$~GeV and $\kappa=1$.
Lower panel: $\Tend=2$~GeV and $\kappa=10^3$.
$Y_\text{eq}$ is also shown in dashed lines.
$\xfo$, $\xeq$, $\xc$ and $\xend$ are also overlaid.
The green horizontal bands correspond to the DM relic abundance, as measured by Planck.
}
\label{fig:yield}
\end{center}
\end{figure}
The first case, characterized by $\Teq\ll\Tfo$ is by far the most studied in the literature.
It corresponds to the scenario where the DM freeze-out happens during radiation domination, and much earlier than the time when $\phi$ decays.
The upper left panel of fig.~\ref{fig:yield} shows the evolution of the DM yield $Y\equiv n/s$ as a function of $x\equiv m/T$, for the benchmark point $m=100$~GeV, $\sv=10^{-11}$~GeV$^{-2}$, $\Tend=7\times10^{-3}$~GeV, $\kappa=10^{-2}$ and $\omega=0$ (point 1 in fig.~\ref{fig:reconstruction}).
The green horizontal band corresponds to the DM relic abundance, as measured by Planck~\cite{Aghanim:2018eyx}.
Here the freeze-out happens as in the standard radiation dominated case, and it is succeeded by a dilution due to the entropy injection produced by the late decay of $\phi$.
Much before the decay of $\phi$, the SM entropy is conserved and therefore the Boltzmann equation~\eqref{eq:cosmo1} can be rewritten as
\begin{equation}\label{eq:cosmo1b}
    \frac{dY}{dx}=-\frac{\sv\,s}{H\,x}\left(Y^2-Y_\text{eq}^2\right),
\end{equation}
where $Y_\text{eq}\equiv n_\text{eq}/s$.
Taking into account that $H\simeq\sqrt{\frac{\rR}{3M_P^2}}=\pi\sqrt{\frac{\gs}{90}}\frac{m^2}{M_P}\frac{1}{x^2}$, eq.~\eqref{eq:cosmo1b} admits the standard approximate solution
\begin{equation}\label{eq:Y1}
    Y_0=\frac{15}{2\pi\sqrt{10\,\gs}}\frac{\xfo}{m\,M_P\,\sv}\,,
\end{equation}
where $Y_0$ corresponds to the DM yield long after the freeze-out, but before the decay of $\phi$.

Additionally, $\xfo\equiv m/\Tfo$ is defined by $\left.n_\text{eq}\sv/H\right|_{x=\xfo}=1$ and given by
\begin{equation}
    \xfo=\ln\left[\frac32\sqrt{\frac{5}{\pi^5\gs}}\,g\,m\,M_P\,\sv\sqrt{\xfo}\right],
\end{equation}
where $g$ is the number of degrees of freedom for DM.
In this first scenario, $\xfo$ is independent on the cosmological parameters $\Tend$, $\kappa$ and $\omega$, because the freeze-out happens in the standard cosmological scenario.
At this point let us emphasize that the obtained DM abundance is much larger than the observed one (as we are assuming that $\sv\ll\sv_0$), and therefore has to be reduced.

The decay of $\phi$ dilutes the DM by injecting entropy to the SM bath.
The dilution factor $D\equiv s(T_2)/s(T_1)=(T_2/T_1)^3$ is defined as the ratio of the SM entropies after and before the decay, and can be estimated as follows.
In the sudden decay approximation of $\phi$, the conservation of the energy density implies
\begin{equation}
    \rR(T_1)+\rp(T_1)=\rR(T_2),
\end{equation}
where $T_1$ and $T_2$ are the temperatures just before and just after $\phi$ decays, respectively.
Taking into account that the scaling of $\rp$ and that $\rp(m)=\kappa\,\rR(m)$, one gets that
\begin{eqnarray}
    D&=&\left(\frac{T_2}{T_1}\right)^3\simeq\left[\kappa\left(\frac{m}{T_2}\right)^{1-3\omega}\right]^\frac{1}{1+\omega}\qquad\text{ for }\omega\neq-1,\label{eq:dil1}\\
    D&=&\left(\frac{T_2}{T_1}\right)^3=\left[1-\kappa\left(\frac{m}{T_2}\right)^4\right]^{-\frac34}\qquad\text{ for }\omega=-1.\label{eq:dil2}
\end{eqnarray}
It can be checked that the choice $T_2=\Tend$ fits well the full numerical solution.

The final DM abundance given by the ratio of eqs.~\eqref{eq:Y1} and~\eqref{eq:dil1} or~\eqref{eq:dil2} has to match the observations by the Planck collaboration~\cite{Aghanim:2018eyx}
\begin{eqnarray}\label{eq:Y1b}
    Y_\text{obs}&=&\frac{Y_0}{D}\simeq\frac{15}{2\pi\sqrt{10\,\gs}}\frac{\xfo}{m\,M_P\,\sv}\left[\frac{1}{\kappa}\left(\frac{\Tend}{m}\right)^{1-3\omega}\right]^\frac{1}{1+\omega}\qquad\text{ for }\omega\neq-1,\\
    Y_\text{obs}&=&\frac{Y_0}{D}=\frac{15}{2\pi\sqrt{10\,\gs}}\frac{\xfo}{m\,M_P\,\sv}\left[1-\kappa\left(\frac{m}{\Tend}\right)^4\right]^\frac34\hspace{1.15cm}\text{ for }\omega=-1,
\end{eqnarray}
where $Y_\text{obs}\times m=\frac{\rho_c\,\Omega_\text{DM} h^2}{s_0\,h^2}\simeq 4\times 10^{-10}$~GeV, $\rho_c$ is the critical energy density of the Universe, and $s_0$ and $\Omega_\text{DM}$ are the entropy density and the DM relic abundance nowadays, respectively.
Previous equations implies that in scenario~\hyperref[sec:1]{1}, in order to reproduce the observed DM abundance $\kappa\propto\Tend^{1-3\omega}$.
In the case where $\omega=0$, $\kappa\propto\Tend$ as observed in fig.~\ref{fig:reconstruction}.

%%%%%%%%%%%%%%%%%%%%%%%%%%%%%%%%%%%%%%%%%%%%%%%%%%%%%%%%%%%%%%%%%%%%%%%%%%%%%%%
\subsubsection[Case 2: $\Tc\ll\Tfo\ll\Teq$]{Case 2: $\boldsymbol{\Tc\ll\Tfo\ll\Teq}$}\label{sec:2}
This case corresponds to the scenario where $\Tc\ll\Tfo\ll\Teq$.
In this regime the Hubble expansion rate is driven by $\rp$.
However, $\phi$ is not yet efficiently decaying into SM radiation, so that $T$ is still inversely proportional to the scale factor.
The upper right panel of fig.~\ref{fig:yield} shows the evolution of the DM yield, for the benchmark point $m=100$~GeV, $\sv=10^{-11}$~GeV$^{-2}$, $\Tend=10^{-1}$~GeV, $\kappa=1$ and $\omega=0$ (point 2 in fig.~\ref{fig:reconstruction}).
Compared to the previous case, the main difference here is the expansion of the Universe.
In fact, now
\begin{equation}
    H\simeq\sqrt{\frac{\rp}{3M_P^2}}=\frac{\pi}{3}\sqrt{\frac{\gs}{10}}\frac{m^2}{M_P}\sqrt{\frac{\kappa}{x^{3(1+\omega)}}}\,,
\end{equation}
and therefore eq.~\eqref{eq:cosmo1b} admits the approximate solutions
\begin{eqnarray}
    Y_0&=&\frac{45}{4\pi}\,\frac{1-\omega}{m\,M_P\,\sv}\sqrt{\frac{\kappa}{10\gs}}\, \xfo^{\frac32(1-\omega)}\hspace{1.55cm}\text{ for }\omega\not=1,\label{eq:Y21}\\
    Y_0&=&\frac{15}{2\pi}\frac{1}{m\,M_P\,\sv}\sqrt{\frac{\kappa}{10\gs}}\,\left[\ln\frac{\xend}{\xfo}\right]^{-1}\qquad\text{ for }\omega=1.\label{eq:Y22}
\end{eqnarray}

\begin{figure}[t]
\begin{center}
\includegraphics[height=0.4\textwidth]{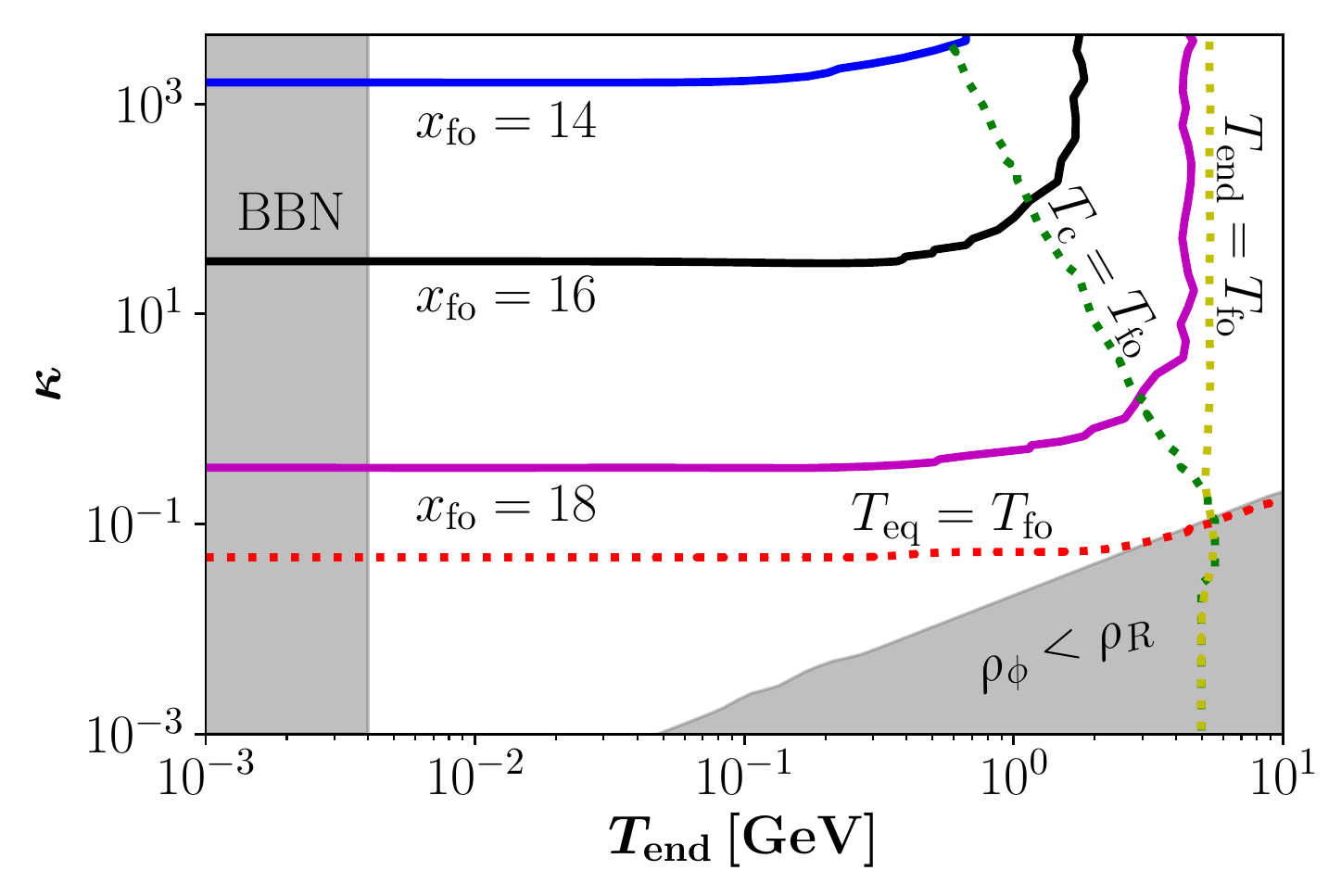}
\caption{Contour lines for the inverse of the temperature at which the DM freeze-out occurs: $\xfo=14$ (blue line), 16 (black line) and 18 (magenta line).
We assumed $\omega=0$, $m=100$~GeV and $\sv=10^{-11}$~GeV$^{-2}$.
The colored bands correspond to $\Tend<\TBBN$ and $\rp<\rR$.
The lines corresponding to $\Tfo=\Teq$, $\Tfo=\Tc$ and $\Tfo=\Tend$ are overlaid.}
\label{fig:FO}
\end{center}
\end{figure}
The DM freeze-out happens at
\begin{equation}\label{eq:xFO2}
    \xfo=\ln\left[\frac32\sqrt{\frac{5}{\pi^5 \,\gs}}\,g\,\frac{m\,M_P\,\sv}{\sqrt{\kappa}}\,\xfo^{\frac32\omega}\right],
\end{equation}
which depends only on $\kappa$.
Figure~\ref{fig:FO} shows contour lines for $\xfo=14$ (blue), 16 (black) and 18 (magenta) in the $[\Tend,\,\kappa]$ plane, assuming $\omega=0$ and numerically solving the full Boltzmann equations.
In the same way as in fig.~\ref{fig:reconstruction}, here we choose $m=100$~GeV and $\sv=10^{-11}$~GeV$^{-2}$.
The left part of the plot in gray, corresponding to $\Tend<4$~MeV, is in tension with BBN.
Additionally, in the lower right corner $\rp$ is always subdominant with respect to radiation, and hence corresponds to the usual case, radiation dominated.
The figure also shows the lines corresponding to $\Tfo=\Teq$, $\Tfo=\Tc$ and $\Tfo=\Tend$.
The $\kappa$ dependence on eq.~\eqref{eq:xFO2}, for $\Tc\ll\Tfo\ll\Teq$, is shown in fig.~\ref{fig:FO} as horizontal lines.

The final DM abundance given by the ratio of eqs.~\eqref{eq:Y21}, \eqref{eq:Y22}, \eqref{eq:dil1} and~\eqref{eq:dil2} is given by
\begin{eqnarray}
    Y_\text{obs}=\frac{Y_0}{D}&=&\frac{45}{4\pi\sqrt{10\gs}}\frac{\sqrt{\kappa}}{m\,M_P\,\sv}\,\xfo^3\left[1-\kappa\left(\frac{m}{\Tend}\right)^4\right]^\frac34\hspace{1.65cm}\text{ for }\omega=-1,\\
    Y_\text{obs}=\frac{Y_0}{D}&\simeq&\frac{45(1-\omega)}{4\pi\sqrt{10\gs}}\frac{\sqrt\kappa}{m\,M_P\,\sv} \,\xfo^{\frac32(1-\omega)}  \left[\frac{1}{\kappa}\left(\frac{\Tend}{m}\right)^{1-3\omega}\right]^\frac{1}{1+\omega}\quad\text{ for }|\omega|\neq1,\quad\label{eq:Y2b}\\
    Y_\text{obs}=\frac{Y_0}{D}&\simeq&\frac{15}{2\pi}\sqrt{\frac{1}{10\,\gs}}\frac{1}{\Tend\,M_P\,\sv}\,\left[\ln\frac{\Tfo}{\Tend}\right]^{-1}\hspace{2.7cm}\text{ for }\omega=1.
\end{eqnarray}
Previous equations imply that in scenario~\hyperref[sec:2]{2}, in order to reproduce the observed DM abundance $\kappa\propto\Tend^{2\frac{1-3\omega}{1-\omega}}$.
In the case of $\omega=0$, $\kappa\propto\Tend^2$ as observed in fig.~\ref{fig:reconstruction}, for $\Tc\ll\Tfo\ll\Teq$.

%%%%%%%%%%%%%%%%%%%%%%%%%%%%%%%%%%%%%%%%%%%%%%%%%%%%%%%%%%%%%%%%%%%%%%%%%%%%%%%
\subsubsection[Case 3: $\Tend\ll\Tfo\ll\Tc$]{Case 3: $\boldsymbol{\Tend\ll\Tfo\ll\Tc}$}\label{sec:3}
This case corresponds to the scenario where $\Tend\ll\Tfo\ll\Tc$.%
\footnote{It is interesting to note that this case is only possible for $\omega\ne-1$; in fact if $\omega=-1$, between $\ac$ and $\aend$ the temperature is independent of the scale factor (see eq.~\eqref{eq:temperature}) and therefore $\Tc=\Tend$.}
In this regime $\rp$ controls both the Hubble expansion rate and the evolution of $\rR$.
The lower panel of fig.~\ref{fig:yield} shows the evolution of the DM yield, for the benchmark point $m=100$~GeV, $\sv=10^{-11}$~GeV$^{-2}$, $\Tend=2$~GeV, $\kappa=10^3$ and $\omega=0$ (point 3 in fig.~\ref{fig:reconstruction}).
As in this case the freeze-out occurs when the $\phi$ is decaying and the SM entropy is not conserved, one can not use anymore the Boltzmann equation~\eqref{eq:cosmo1b}.
Instead, eq.~\eqref{eq:cosmo1} can be rewritten as
\begin{equation}\label{eq:BEcase3}
    \frac{dN}{da}=-\frac{\sv}{H\,a^4}\left(N^2-N_\text{eq}^2\right),
\end{equation}
where $N\equiv n\times a^3$ and similarly $N_\text{eq}\equiv n_\text{eq}\times a^3$.
Taking into account that
\begin{equation}
H(a)\simeq\sqrt{\frac{\rp(a)}{3M_P^2}}=\sqrt{\frac{\rp(a_0)}{3M_P^2}}\left(\frac{a_0}{a}\right)^{\frac32(1+\omega)}=\frac{\pi}{3}\sqrt{\kappa\,\frac{\gs}{10}}\,\frac{m^2}{M_P}\left(\frac{a_0}{a}\right)^{\frac32(1+\omega)},
\end{equation}
and choosing the scale factor such that $a_0\equiv a(T=m)=1$, eq.~\eqref{eq:BEcase3} admits the approximate solution
\begin{eqnarray}\label{eq:N0}
    N_0&=&\frac{(1-\omega)\,\pi}{2}\sqrt{\kappa\,\frac{\gs}{10}}\,\frac{m^2}{M_P\sv}\left(\frac{\afo}{a_0}\right)^{\frac32(1-\omega)}\qquad\text{ for }\omega\ne1,\\
    N_0&=&\frac{\pi}{3}\sqrt{\kappa\,\frac{\gs}{10}}\,\frac{m^2}{M_P\sv}\,\left(\ln\frac{\aend}{\afo}\right)^{-1}\hspace{2cm}\text{ for }\omega=1,
\end{eqnarray}
where $N_0$ corresponds to the value of $N$ well after the DM freeze-out.

Within the sudden decay approximation and using eq.~\eqref{eq:temperature}, the value of the critical temperature $\Tc$ and the scale factors at $T=\Tfo$ and $T=\Tend$ can be estimated as
\begin{eqnarray}
    \Tc&=&\left(\kappa\,m^{1-3\omega}\,\Tend^4\right)^{\frac{1}{5-3\omega}},\label{eq:Tc}\\
    \afo&=&a_0\,m\left(\frac{\Tc^{5-3\omega}}{\Tfo^8}\right)^\frac{1}{3(1+\omega)}=a_0\,\left[\kappa\left(\frac{m\,\Tend}{\Tfo^2}\right)^4\right]^\frac{1}{3(1+\omega)},\label{eq:afo3}\\
    \aend&=&a_0\,m\left(\frac{\Tc^{5-3\omega}}{\Tend^8}\right)^\frac{1}{3(1+\omega)}=a_0\,\left[\kappa\left(\frac{m}{\Tend}\right)^4\right]^\frac{1}{3(1+\omega)}.\label{eq:aend3}
\end{eqnarray}

The final DM yield $Y_0$ is related to $N_0$ via the factor $s\times a^3$, which after the decay of $\phi$ can be written as
\begin{equation}
    s\, a^3=\frac{2\pi^2}{45}\gs\left(\Tend\,\aend\right)^3=\frac{2\pi^2}{45}\gs\left[\kappa\frac{m^4}{\Tend^{1-3\omega}}\right]^\frac{1}{1+\omega},
\end{equation}
implying that
\begin{eqnarray}
    Y_\text{obs}&=&\frac{N_0}{s\,a^3}=\frac{45(1-\omega)}{4\pi}\sqrt{\frac{1}{10\gs}}
    \frac{1}{M_P\sv}\left[\Tfo^{4(\omega-1)}\,\Tend^{3-5\omega}\right]^\frac{1}{1+\omega}\quad\text{ for }\omega\ne1,\label{eq:Y3}\\
    Y_\text{obs}&=&\frac{N_0}{s\,a^3}=\frac{45}{8\pi}\sqrt{\frac{1}{10\gs}}\frac{1}{\Tend\,M_P\,\sv}\,\left(\ln\frac{\Tfo}{\Tend}\right)^{-1}\hspace{1.8cm}\text{ for }\omega=1,
\end{eqnarray}
which is independent of $\kappa$, as expected from fig.~\ref{fig:reconstruction}.

In order to estimate the temperature at which the DM freeze-out happens, let us first examine how $\rp$ scales.
The evolution of $\rp$ has to be divided in two regimes (after and before $a=\ac$), because of the two different dependences on $T$:
\begin{equation}\label{eq:rho3b}
    \rp(a)=\rp(a_0)\left(\frac{a_0}{a}\right)^{3(1+\omega)}=\rp(a_0)\left(\frac{a_0}{a_c}\frac{a_c}{a}\right)^{3(1+\omega)}.
\end{equation}
Using eqs.~\eqref{eq:temperature} and~\eqref{eq:Tc}, eq.~\eqref{eq:rho3b} can be rewritten as
\begin{equation}
    \rp(T)=\rp(m)\left(\frac{T_c}{m}\right)^{3(1+\omega)}\left(\frac{T}{T_c}\right)^8=\frac{\pi^2\,\gs}{30}\kappa\,m^{1-3\omega}\frac{T^8}{\Tc^{5-3\omega}}=\frac{\pi^2\,\gs}{30}\left(\frac{T^2}{\Tend}\right)^4,
\end{equation}
which turns out to be $\kappa$-independent.
Therefore the DM freeze-out happens at
\begin{equation}\label{eq:xFO3}
    \xfo=\ln\left[\frac32\sqrt{\frac{5}{\pi^5\,\gs}}\,g\, \frac{M_P\,\sv\,\Tend^2}{m}\,\xfo^\frac52\right],
\end{equation}
or equivalently
\begin{equation}\label{eq:xFO3b}
    \xfo=-\frac52\,W_{-1}\left[-\frac25\left(\frac32\sqrt{\frac{5}{\pi^5\,\gs}}\,g\, \frac{M_P\,\sv\,\Tend^2}{m}\right)^{-\frac25}\right],
\end{equation}
where $W_{-1}$ is the $-1$ branch of the Lambert $W$ function, and which is again independent on $\kappa$.
Figure~\ref{fig:FO} shows the $\Tend$ dependence of $\xfo$ as vertical lines.

%%%%%%%%%%%%%%%%%%%%%%%%%%%%%%%%%%%%%%%%%%%%%%%%%%%%%%%%%%%%%%%%%%%%%%%%%%%%%%%
\subsubsection[Case 4: $\Tfo\ll\Tend$]{Case 4: $\boldsymbol{\Tfo\ll\Tend}$}\label{sec:4}
This case corresponds to the scenario where $\Tfo\ll\Tend$.
In this regime the non-standard cosmology has no effect on the final DM relic abundance, due to the fact that $\phi$ decays at a very high temperature, while the DM is still in chemical equilibrium with the SM thermal bath.\\

%%%%%%%%%%%%%%%%%%%%%%%%%%%%%%%%%%%%%%%%%%%%%%%%%%%%%%%%%%%%%%%%%%%%%%%%%%%%%%%
\subsection{Varying the Particle Physics Parameters}
Up to now we have studied the possibilities for reconstructing non-standard cosmologies after a DM detection assuming some given particle physics benchmarks.
In this section we study the reconstruction prospects using different benchmarks both for the DM properties ($m$ and $\sv$), and the equation of state $\omega$ of $\phi$.

\begin{figure}[t]
\begin{center}
\includegraphics[height=0.33\textwidth]{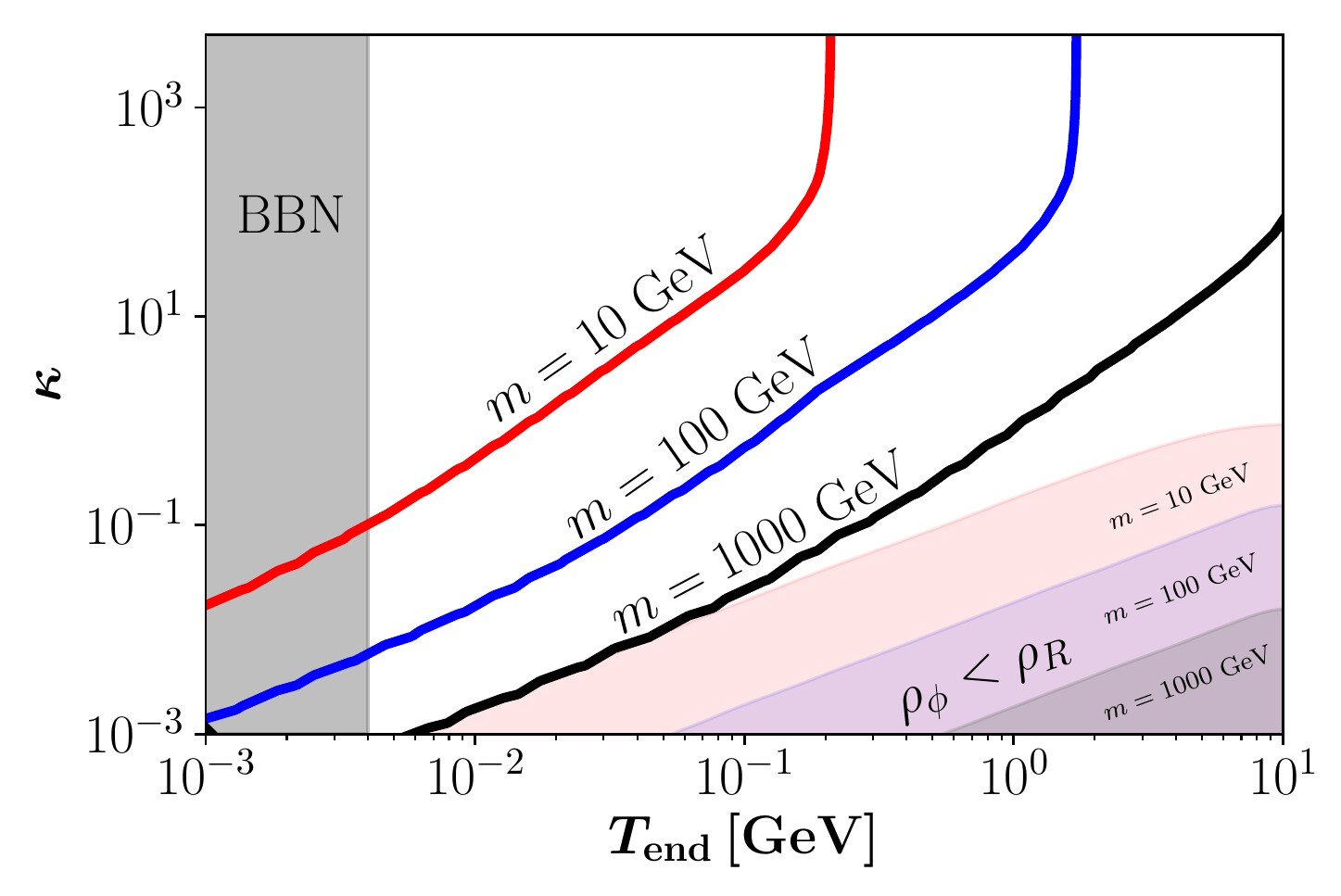}
\includegraphics[height=0.33\textwidth]{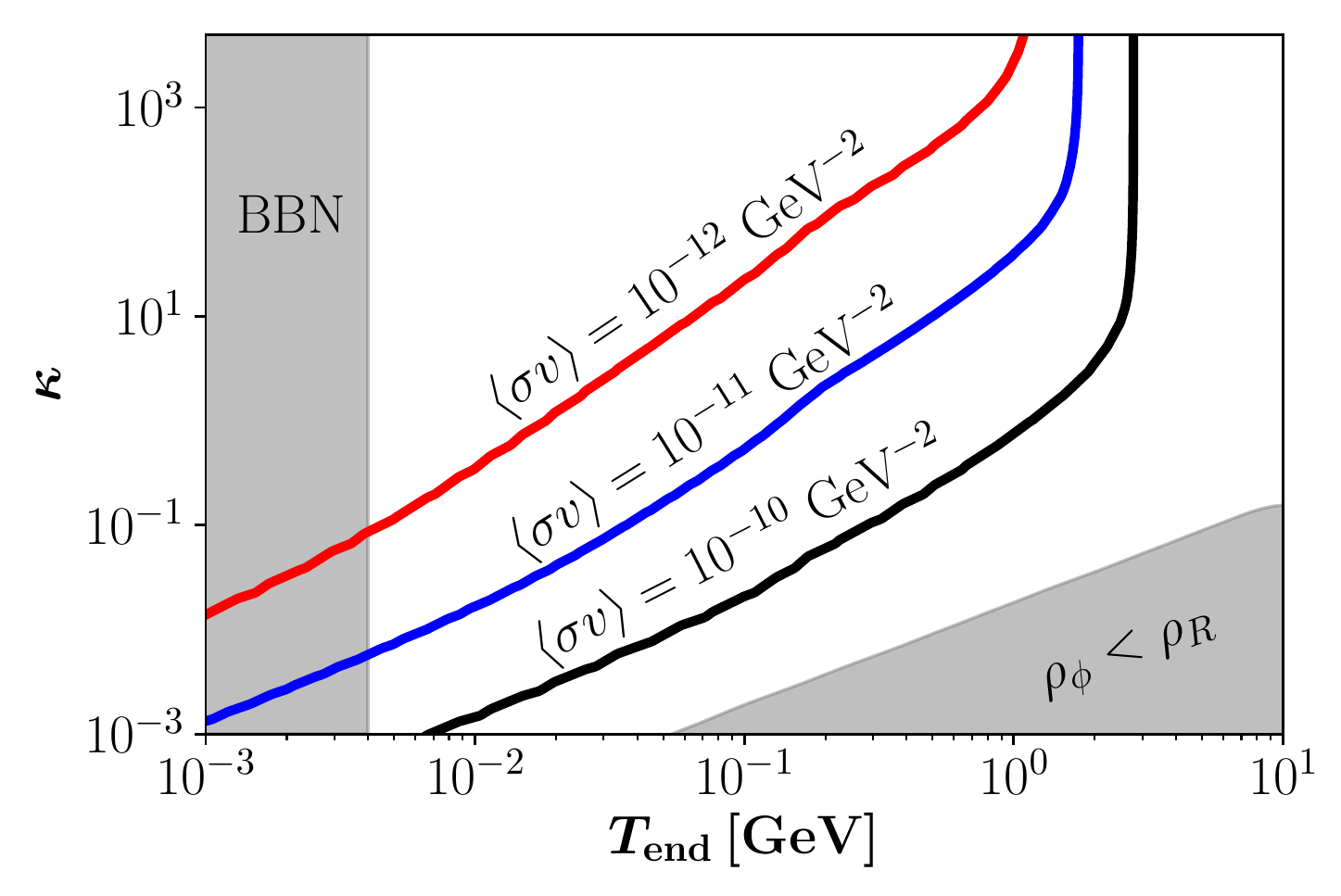}
\caption{Parameter space generating the observed DM abundance via the WIMP mechanism with non-standard cosmologies, assuming $\omega=0$ and different particle physics parameters.
Left panel: $\sv=10^{-11}$~GeV$^{-2}$ and $m=10$~GeV (red), 100~GeV (blue), and 1~TeV (black). The colored bands present the $\rp<\rR$ areas for $m=10$~GeV (light red), 100~GeV (purple), and 1~TeV (gray).
Right panel:  $m=100$~GeV and $\sv=10^{-12}$~GeV$^{-2}$ (red), $10^{-11}$~GeV$^{-2}$ (blue), and $10^{-10}$~GeV$^{-2}$ (black), while the gray right lower band correspond to $\rp<\rR$.
The gray bands on the left correspond to $\Tend<\TBBN$.}
\label{fig:reconstructions}
\end{center}
\end{figure}
Figure~\ref{fig:reconstructions} shows the parameter space generating the observed DM abundance via the WIMP mechanism with non-standard cosmologies assuming $\omega=0$ and different particle physics parameters.
The left panel corresponds to $\sv=10^{-11}$~GeV$^{-2}$ and DM masses: $m=10$~GeV (red), 100~GeV (blue) and 1~TeV (black).
Notice that in this case, the colored bands corresponding to $\rp<\rR$ are different for each mass, because $\kappa$ is defined at different scales ($T=m$).
Additionally, the right panel depicts the cases where $m=100$~GeV and $\sv=10^{-12}$~GeV$^{-2}$ (red), $10^{-11}$~GeV$^{-2}$ (blue) and $10^{-10}$~GeV$^{-2}$ (black).
The gray bands correspond to $\Tend<\TBBN$ and $\rp<\rR$.
The behavior of the lines can be understood analytically.
On the one hand, for low values of $\kappa$, we are in case~\hyperref[sec:1]{1} where the DM relic density scales like $\frac{\Tend}{\sv\,\kappa\,m}$ (up to a mild logarithmic dependence coming from $\xfo$), see eq.~\eqref{eq:Y1b}.
Therefore, an increase of the DM mass or $\sv$ decreases the final DM yield.
This effect can be compensated by reducing the dilution factor $D$ by either a rise of $\Tend$ or a decrease of $\kappa$.
On the other hand, for high values of $\kappa$ we are in case~\hyperref[sec:3]{3}, where the DM relic density scales like $\frac{\Tend^3}{\sv\,m^4}$ (again up to a mild logarithmic dependence coming from $\xfo$), see eq.~\eqref{eq:Y3}.
As pointed out previously, this scenario is independent of $\kappa$.
Therefore, an increase of either $m$ or $\sv$ has to be compensated by a rise of $\Tend$.

\begin{figure}[t]
\begin{center}
\includegraphics[height=0.33\textwidth]{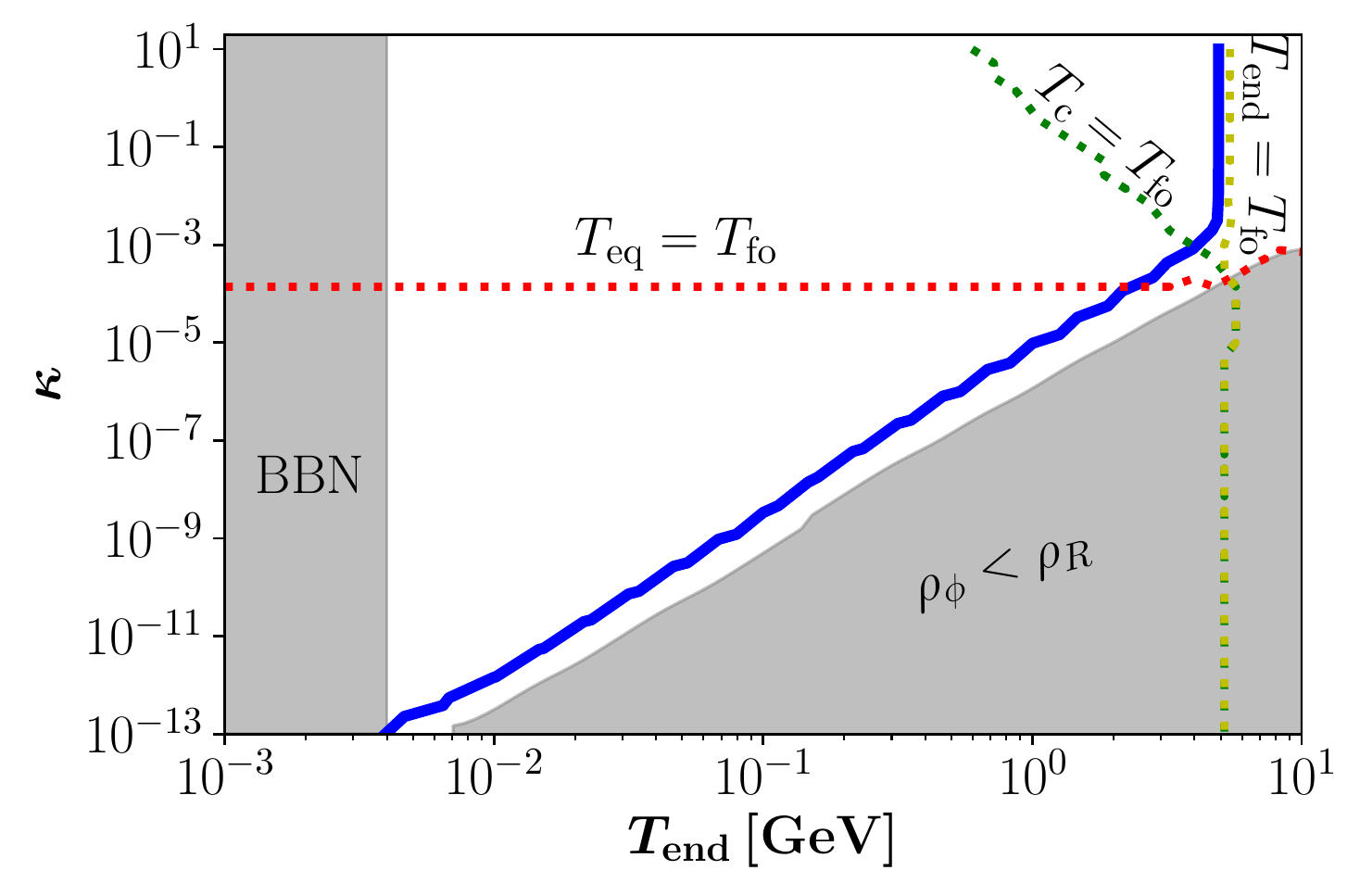}
\includegraphics[height=0.33\textwidth]{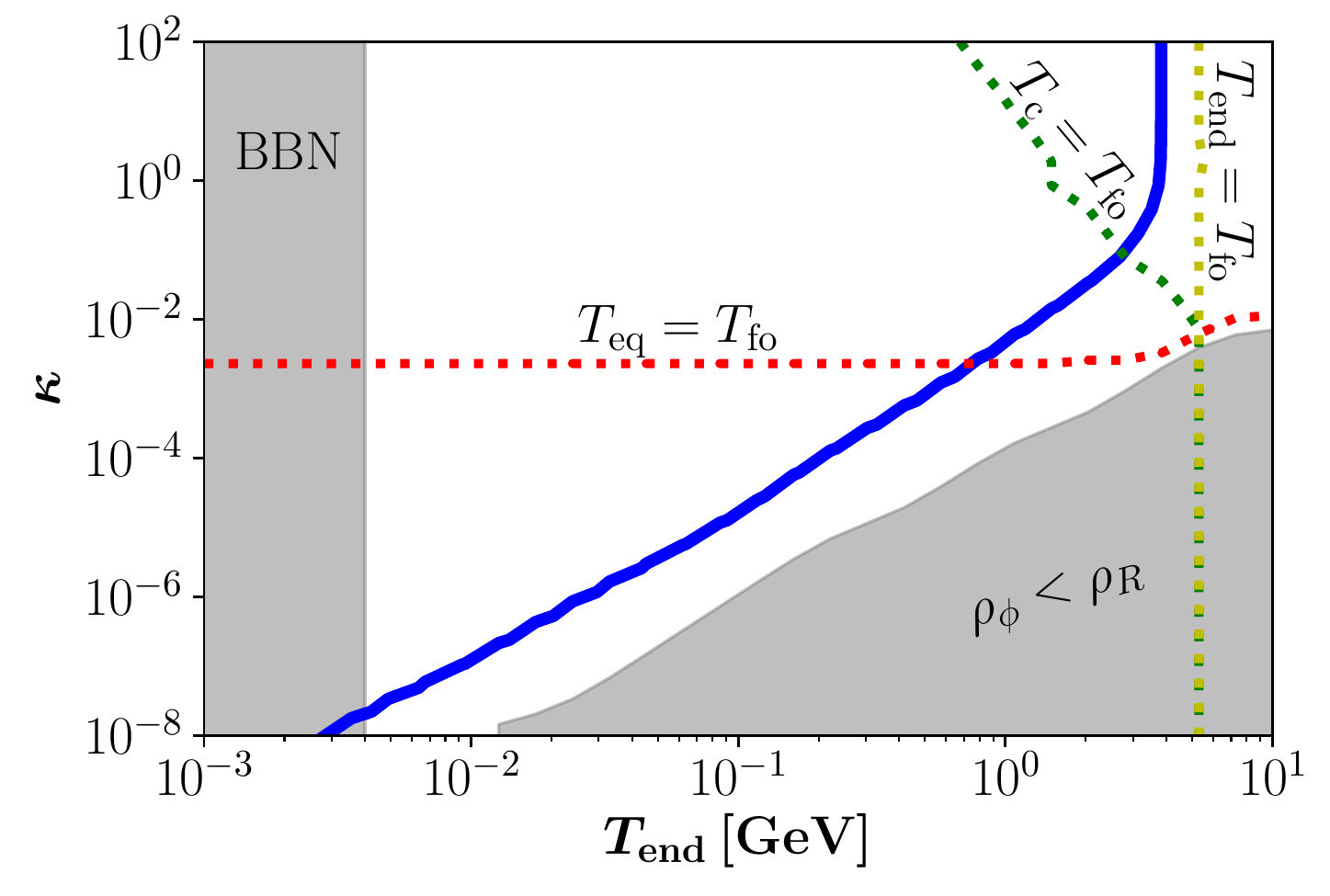}
\includegraphics[height=0.33\textwidth]{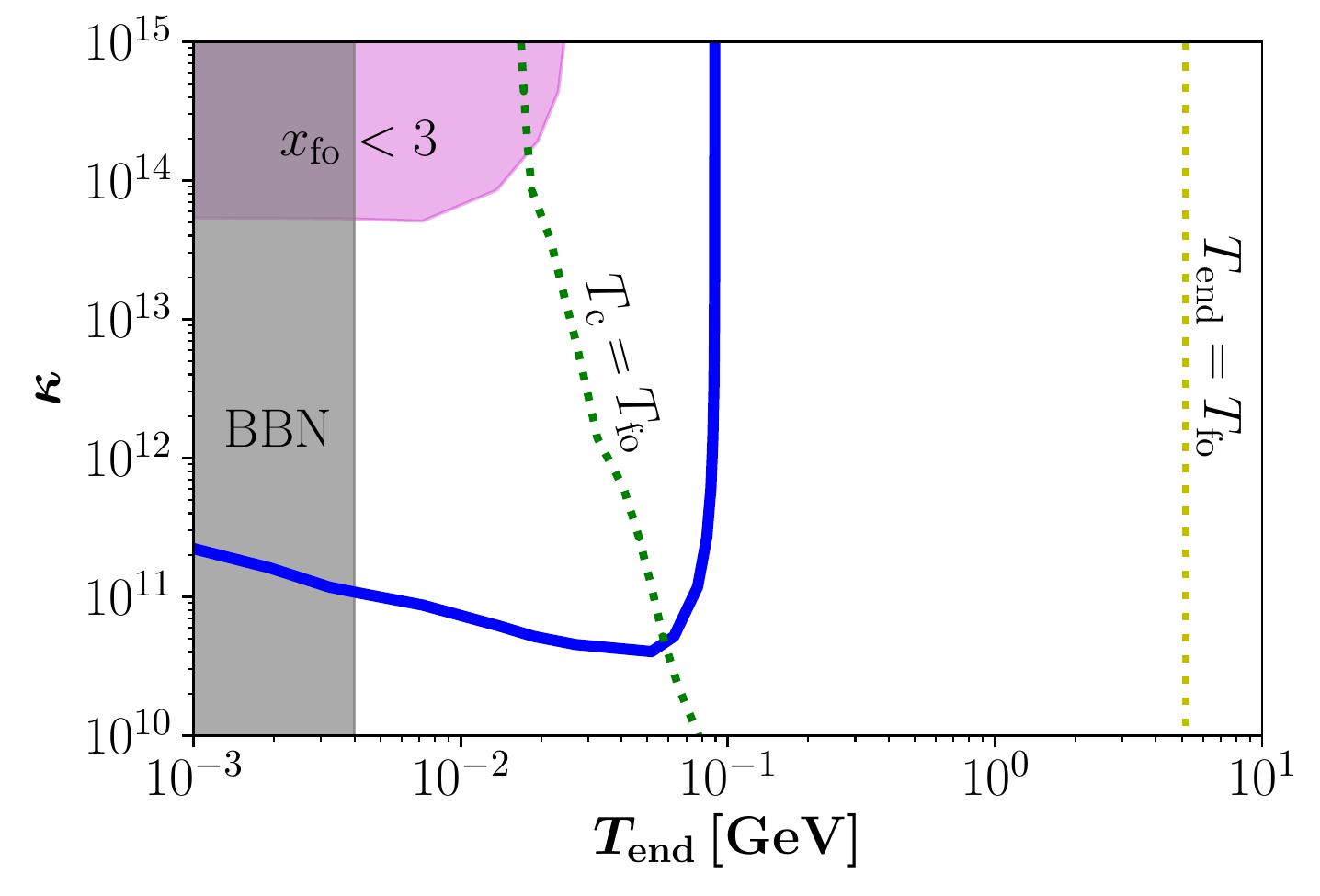}
\caption{Parameter space generating the observed DM abundance via the WIMP mechanism with non-standard cosmologies, assuming $\omega=-2/3$ (upper left panel), $\omega=-1/3$ (upper right panel) and $\omega=+2/5$ (lower panel).
For the particle physics benchmark we have taken $m=100$~GeV and $\sv=10^{-11}$~GeV$^{-2}$.
The gray bands correspond to $\Tend<\TBBN$ and $\rp<\rR$, while the fuchsia region for the $\omega=+2/5$ case shows the (semi-)relativistic freeze-out.
The lines corresponding to $\Tfo=\Teq$, $\Tfo=\Tc$ and $\Tfo=\Tend$ are overlaid.
}
\label{fig:reconstructionomega}
\end{center}
\end{figure}
Figure~\ref{fig:reconstructionomega} presents in blue the parameter space compatible with the observed DM abundance via the WIMP mechanism with non-standard cosmologies, in the plane $[\Tend,\,\kappa]$.
We assumed $\omega=-2/3$ (upper left panel), $\omega=-1/3$ (upper right panel) and $\omega=+2/5$ (lower panel).
For the particle physics benchmark we have chosen $m=100$~GeV and $\sv=10^{-11}$~GeV$^{-2}$.
The gray bands correspond to $\Tend<\TBBN$ and $\rp<\rR$, while the fuchsia region for the $\omega=+2/5$ case shows the (semi-)relativistic freeze-out, i.e. $\xfo<3$.
The lines corresponding to $\Tfo=\Teq$, $\Tfo=\Tc$ and $\Tfo=\Tend$ are overlaid.\\
On the left upper panel $\omega=-2/3$ corresponds to a $\rp$ that scales like $a^{-1}$.
This implies that it gets diluted much slower than matter, and thus naturally dominates the total energy density of the Universe, even if its initial density is suppressed.
One can therefore explore much lower values for $\kappa$, compared to the case $\omega=0$ in fig.~\ref{fig:reconstruction}, without violating the BBN bound.
As expected from the analytical estimations, in the regions $\Teq\ll\Tfo$ (case~\hyperref[sec:1]{1}) and $\Tc\ll\Tfo\ll\Teq$ (case~\hyperref[sec:2]{2}), $\kappa$ scales like $\Tend^3$ and $\Tend^{18/5}$, respectively.
 Also, when $\Tend\ll\Tfo\ll\Tc$ (case~\hyperref[sec:3]{3}) the DM yield is independent on $\kappa$.
In the right lower corner the $\rp$ is always subdominant with respect to radiation, and hence corresponds to the usual case, radiation dominated.\\
Similarly, on the right upper panel $\omega=-1/3$ corresponds to a $\rp$ that scales like $a^{-2}$.
In the regions $\Teq\ll\Tfo$ (case~\hyperref[sec:1]{1}) and $\Tc\ll\Tfo\ll\Teq$ (case~\hyperref[sec:2]{2}), $\kappa$ scales like $\Tend^2$ and $\Tend^3$, respectively.
 Also, when $\Tend\ll\Tfo\ll\Tc$ (case~\hyperref[sec:3]{3}) the DM yield is independent on $\kappa$.\\
On contrary, the lower panel corresponds to $\omega=+2/5$ and hence $\rp\propto a^{-21/5}$.
The $\phi$ energy density gets diluted much faster than matter (and radiation) and therefore very large values for $\kappa$ are needed to compensate.
In turn, large values of $\kappa$ boost the Hubble expansion rate implying a much earlier freeze-out.
The upper left region corresponds to $\xfo<3$, yielding a (semi-)relativistic freeze-out which is incompatible with our approximations.
In this case with $\omega>1/3$ and $\kappa\gg1$, $\Teq$ is not defined, as $\rp=\rR$ only happens when $\phi$ decays, implying that the case~\hyperref[sec:1]{1} is never realized.
Additionally, in the case~\hyperref[sec:2]{2}, $\kappa\propto \Tend^{-2/3}$ and in the case~\hyperref[sec:3]{3} the DM yield is again independent of $\kappa$.

%%%%%%%%%%%%%%%%%%%%%%%%%%%%%%%%%%%%%%%%%%%%%%%%%%%%%%%%%%%%%%%
\subsection{Varying the Non-standard Cosmological Parameters}

In this section we study the impact of the non-standard cosmology on the particle physics parameter space $[m,\,\sv]$.

\begin{figure}[t]
\begin{center}
\includegraphics[height=0.32\textwidth]{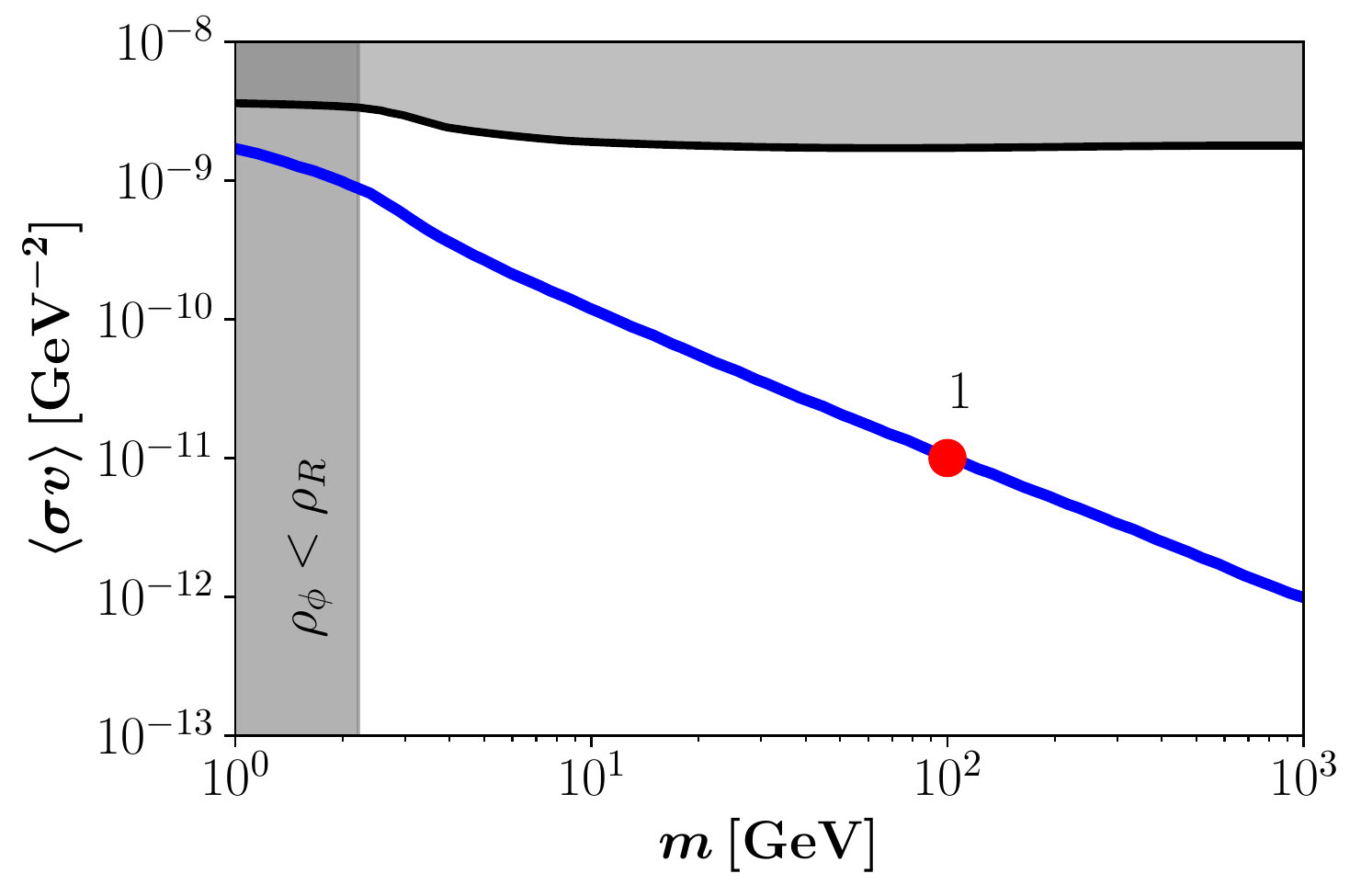}
\includegraphics[height=0.32\textwidth]{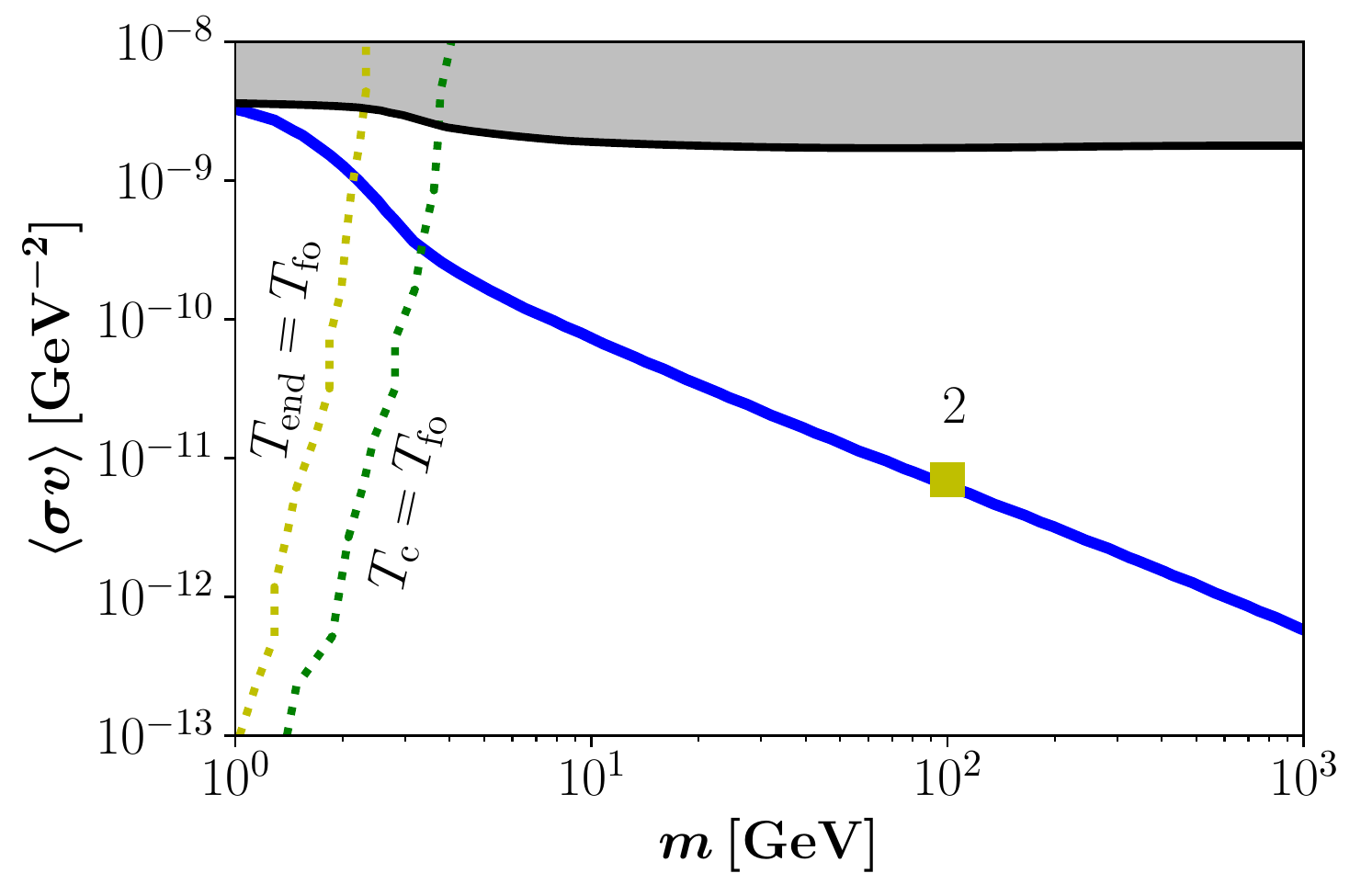}
\includegraphics[height=0.32\textwidth]{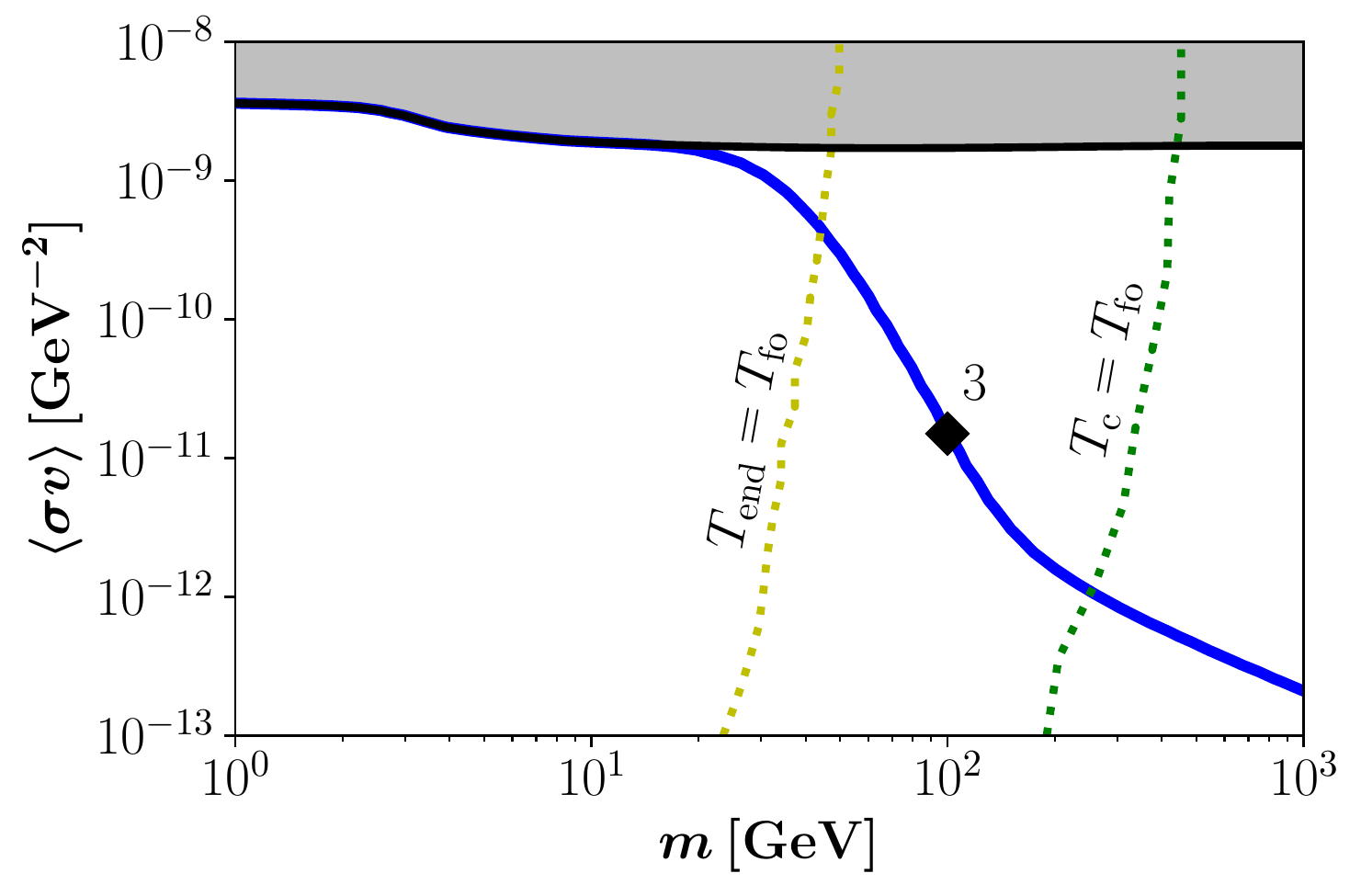}
\caption{Parameter space (in blue) that reproduces the observed DM abundance via the WIMP mechanism, assuming $\Tend=7 \times 10^{-3}$~GeV and $\kappa=10^{-2}$ (upper left panel), $\Tend=10^{-1}$~GeV and $\kappa=1$ (upper right panel), and $\Tend=2$~GeV and $\kappa=10^3$ (lower panel), for $\omega=0$.
The red, yellow and black markers correspond to the benchmark points shown in fig.~\ref{fig:reconstruction}.
The black line ($\sv=\sv_0$) shows the cross sections needed to have a WIMP production with standard cosmology.
Larger cross sections (in gray) are incompatible with the WIMP mechanism, even in the cases of non-standard cosmologies.
The lines corresponding to $\Tfo=\Tc$ and $\Tfo=\Tend$ are overlaid.
}
\label{fig:pre_sv-m}
\end{center}
\end{figure}
Figure~\ref{fig:pre_sv-m} shows in blue the particle physics parameter space $[m,\,\sv]$ that gives rise to the observed DM abundance, for fixed non-standard cosmologies, $\Tend=7\times10^{-3}$~GeV and $\kappa=10^{-2}$ (upper left panel), $\Tend=10^{-1}$~GeV and $\kappa=1$ (upper right panel), and $\Tend=2$~GeV and $\kappa=10^3$ (lower panel), assuming $\omega=0$.
The black line, for which $\sv=\sv_0$, shows the thermally averaged cross sections needed to have a WIMP production with standard cosmology.
The small variations are due to the changes of the number of relativistic degrees of freedom $\gs$ and $\gss$~\cite{Steigman:2012nb}.
Larger cross sections (in gray) are incompatible with the WIMP mechanism, even in the cases of non-standard cosmologies.
The red, yellow and black markers correspond to the benchmark points shown in fig.~\ref{fig:reconstruction}.
The dotted lines corresponding to $\Tfo=\Tc$ (green) and $\Tfo=\Tend$ (yellow) are overlaid.\\
In the upper left panel ($\Tend=7\times10^{-3}$~GeV and $\kappa=10^{-2}$) the gray band on the left corresponds to $\rR>\rp$, i.e. the limit of standard cosmology.
That panel correspond to the case~\hyperref[sec:1]{1}, where $\Teq<\Tfo$.
In order to keep a constant DM relic abundance, in eq.~\eqref{eq:Y1b} $m\times Y_\text{obs}$ has to stay constant as well.
That implies that $\sv\propto m^\frac{3\omega-1}{1+\omega}$, which translates to $\sv\propto 1/m$ for $\omega=0$.\\
A similar behavior appears in the upper right and lower panels, when $\Tc\ll\Tfo\ll\Teq$ (case~\hyperref[sec:2]{2}), which corresponds to high masses, to the right of the dotted green lines:
$\sv\propto m^\frac{3\omega-1}{1+\omega}$, see eq.~\eqref{eq:Y2b}.
For $\omega=0$ it translates to $\sv\propto m^{-1}$.
However, for intermediate masses, between the yellow and the green lines (i.e. for $\Tend\ll\Tfo\ll\Tc$), case~\hyperref[sec:3]{3} happens.
In that scenario, for keeping constant the DM relic abundance $\sv\propto m^\frac{5\omega-3}{1+\omega}$, which for $\omega=0$ implies $\sv\propto m^{-3}$, see eq.~\eqref{eq:Y3}.
Finally, for low masses, to the left of the dotted yellow lines $\Tfo<\Tend$, and therefore the cross section needed to have a successful WIMP DM production is the usual $\sv_0$, characteristic of the standard cosmology (case~\hyperref[sec:4]{4}).

\begin{figure}[t]
\begin{center}
\includegraphics[height=0.4\textwidth]{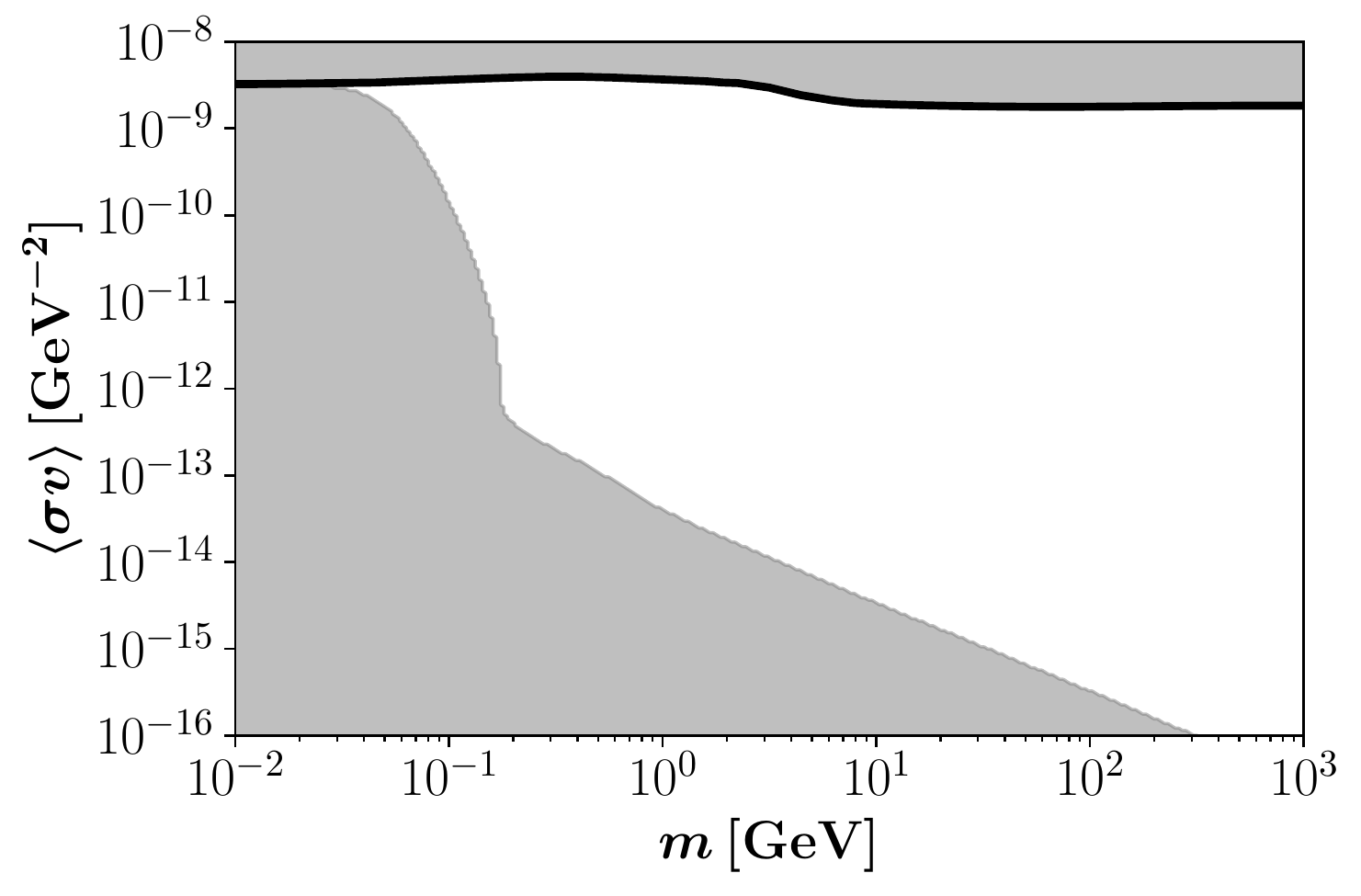}
\caption{Parameter space (in white) that could reproduce the observed DM abundance via the WIMP mechanism with non-standard cosmologies, assuming $\omega=0$.
For completeness we also show the standard cosmological scenario with $\sv=\sv_0$ with a thick black line.
}
\label{fig:sv-m}
\end{center}
\end{figure}
Figure~\ref{fig:sv-m} depicts in white the particle physics parameter space $[m,\,\sv]$ that could reproduce the observed DM abundance via the WIMP mechanism with non-standard cosmologies, assuming $\omega=0$.
The case $\sv=\sv_0=$ few$\times 10^{-9}$~GeV$^{-2}$ giving rise to the simplest WIMP mechanism with the standard cosmology is also shown with a thick black line.
The gray regions show the areas where different cosmologies can not conciliate a DM with mass $m$ and cross-section $\sv$ with the WIMP paradigm.
On the one hand, the case $\omega=0$ can only dilute the DM abundance, which means that cross sections higher than $\sv_0$ can not become viable (upper gray region).
On the other hand, a large part of the parameter space $\sv<\sv_0$ (in white) becomes compatible with the WIMP mechanism.
However, the observed DM relic abundance can not be reproduced for arbitrarily small values for $\sv$ without reaching the (semi-)relativistic freeze-out limit, $\xfo<3$ (lower gray region).
For masses $m\gtrsim 300$~MeV, the bound corresponds to the case~\hyperref[sec:2]{2}, and can be analytically understood by the use of eqs.~\eqref{eq:xFO2} and~\eqref{eq:Y2b}:

\begin{equation}
    \sv=  \left[ \frac{45\,(1-\omega)}{4\pi\,\sqrt{10g_*}} \frac{1}{M_P\, Y_\text{obs}\,m} \right] ^\frac{1+\omega}{2} \left[ \frac{2}{3g} \sqrt{\frac{\pi^5\,g_*}{5}} \frac{\xfo^\frac32\,e^{\xfo}}{M_P} \right] ^\frac{1-\omega}{2} \frac{\Tend^\frac{1-3\omega}{2}}{m^{1-2\omega}},
\end{equation}
that for $\omega=0$ gives $\sv \propto m^{-1}$, and takes the minimum allowed value when $\xfo=3$ and $\Tend=\TBBN$.
Likewise, for 10~MeV $\lesssim m\lesssim 300$~MeV the bound comes from case~\hyperref[sec:3]{3}.
Equation~\eqref{eq:Y3} can be rewritten as
\begin{equation}
    \sv= \frac{45}{4\pi\,\sqrt{10g_*}} \frac{1-\omega}{M_P\, Y_\text{obs}\,m}\, \xfo^{4\frac{1-\omega}{1+\omega}} \left( \frac{\Tend}{m} \right) ^\frac{3-5\omega}{1+\omega},
\end{equation}
which again represents a lower bound when taking $\Tend=\TBBN$ and $\xfo$ as given in eq.~\eqref{eq:xFO3b}.
From fig.~\ref{fig:sv-m} one can see that DM lighter than $\sim 30$~MeV can only be produced in the standard cosmological scenario with $\sv=\sv_0$.
For those masses, our non-standard cosmological setup with $\omega=0$ can not conciliate smaller cross sections with the WIMP mechanism.

Before closing this section we would like to emphasize that in this work the thermally averaged cross section $\sv$ is evaluated at freeze-out, i.e. when DM velocity is $v\simeq 1/3$.
There are bounds at much lower redshifts coming from CMB~\cite{Slatyer:2015jla}, the galactic center~\cite{TheFermi-LAT:2017vmf} and dwarfs galaxies~\cite{Fermi-LAT:2016uux}.
However these bounds depend on the velocity scaling of $\sv$ and, in this model independent approach, can not be applied directly.

%%%%%%%%%%%%%%%%%%%%%%%%%%%%%%%%%%%%%%%%%%%%%%%%%%%%%%%%%%%%%%%%%%%%%%%%%%%%%%%
%%%%%%%%%%%%%%%%%%%%%%%%%%%%%%%%%%%%%%%%%%%%%%%%%%%%%%%%%%%%%%%%%%%%%%%%%%%%%%%
%%%%%%%%%%%%%%%%%%%%%%%%%%%%%%%%%%%%%%%%%%%%%%%%%%%%%%%%%%%%%%%%%%%%%%%%%%%%%%%
\section{Conclusions}\label{sec:Conclusion}
Despite the large amount of searches over the past decades, dark matter (DM) has not been found.
In particular, scenarios where DM is a weakly interacting massive particle (WIMP) have received by far the biggest attention both theoretically and experimentally, but unfortunately there is no overwhelming evidence of WIMP DM.
A simple reason for this might be that the cosmological history was non-standard at early times, which affects the typical DM interaction rates, reducing the naively expected annihilation cross sections.

In this paper we considered the production of WIMP DM in the early Universe following a particle physics model independent way, where the DM dynamics is fully parametrized by its mass $m$ and its total thermally averaged annihilation cross section $\sv$.
Additionally, we studied scenarios where for some period the expansion of the Universe was governed by a component $\phi$ with an effective equation of state $\omega=p_\phi/\rp$, where $p_\phi$ is its pressure and $\rp$ its energy density.

Once DM is discovered and its particle physics properties have been reconstructed, a major question rises concerning the DM production mechanism.
If the inferred value of $\sv$ is in the ballpark of few$\times 10^{-26}$~cm$^3$/s, the simpler freeze-out mechanism with a standard cosmology will be strongly favored.
However, if that turns out not to be the case, one can either look for different DM production mechanisms or for alternative cosmological scenarios.
The latter option was pursued in this study.

A detailed analysis was performed both numerically and analytically, by solving the system of coupled Boltzmann equations.
Different regimes have been found, characterized by the temperature at which the DM freeze-out happens compared to the proper scales of the non-standard cosmology.
We studied the effects of varying both the particles physics and the non-standard cosmological parameters, and found the parameter space that was compatible with the observed DM abundance via the WIMP paradigm.

We found that large regions of the particle physics parameter space can be reconciled with the WIMP paradigm in the case of non-standard cosmologies for DM heavier than $\sim 30$~MeV.
An effect on the genesis of lighter WIMP DM without modifying the usual BBN dynamics is not possible within our approach.
On the contrary, heavy DM WIMP can be compatible with the WIMP mechanism and cross sections much smaller than the canonical $\sv_0=$ few$\times 10^{-9}$~GeV$^{-2}$.
In particular, for TeV DM one can go to values as small as $\sv\simeq 3\times10^{-17}$~GeV$^{-2}\sim 3\times 10^{-34}$~cm$^3$/s.

%%%%%%%%%%%%%%%%%%%%%%%%%%%%%%%%%%%%%%%%%%%%%%%%%%%%%%%%%%%%%%%%%%%%%%%%%%%%%%%
%%%%%%%%%%%%%%%%%%%%%%%%%%%%%%%%%%%%%%%%%%%%%%%%%%%%%%%%%%%%%%%%%%%%%%%%%%%%%%%
%%%%%%%%%%%%%%%%%%%%%%%%%%%%%%%%%%%%%%%%%%%%%%%%%%%%%%%%%%%%%%%%%%%%%%%%%%%%%%%
\acknowledgments
We would like to thank Fazlollah Hajkarim and Sergio Palomares-Ruiz for valuable discussions.
NB is partially supported by Spanish MINECO under Grant FPA2017-84543-P. 
AH acknowledges the Joven Investigador program by Universidad Antonio Nariño.
This project has received funding from the European Union's Horizon 2020 research and innovation programme under the Marie Sklodowska-Curie grant agreements 674896 and 690575, and from Universidad Antonio Nariño grants 2017239, 2018204, 2019101 and 2019248.
CM is supported by CONICYT-
PCHA/Doctorado Nacional/2018-21180309. PA and CM are supported by FONDECYT Project 1161150.
This research made use of IPython~\cite{Perez:2007emg}, Matplotlib~\cite{Hunter:2007ouj} and SciPy~\cite{SciPy}.

%%%%%%%%%%%%%%%%%%%%%%%%%%%%%%%%%%%%%%%%%%%%%%%%%%%%%%%%%%%%%%%%%%%%%%%%%%%%%%%%%%%%%%%%%%%%%%%%%%%%%%%%%%%%%%%%%%%%%%%%%%%%%%%%%%%%%%%%%%%%%%%%%%%%%%%%%%%%%%%%%%%%%%%%

\bibliography{biblio}

\providecommand{\href}[2]{#2}\begingroup\raggedright\begin{thebibliography}{100}

\bibitem{Aghanim:2018eyx}
{\scshape Planck} collaboration, \emph{{Planck 2018 results. VI. Cosmological
  parameters}},  \href{https://arxiv.org/abs/1807.06209}{{\ttfamily
  1807.06209}}.

\bibitem{Bertone:2004pz}
G.~Bertone, D.~Hooper and J.~Silk, \emph{{Particle dark matter: Evidence,
  candidates and constraints}},
  \href{https://doi.org/10.1016/j.physrep.2004.08.031}{\emph{Phys. Rept.}
  {\bfseries 405} (2005) 279}
  [\href{https://arxiv.org/abs/hep-ph/0404175}{{\ttfamily hep-ph/0404175}}].

\bibitem{Arcadi:2017kky}
G.~Arcadi, M.~Dutra, P.~Ghosh, M.~Lindner, Y.~Mambrini, M.~Pierre et~al.,
  \emph{{The waning of the WIMP? A review of models, searches, and
  constraints}},
  \href{https://doi.org/10.1140/epjc/s10052-018-5662-y}{\emph{Eur. Phys. J.}
  {\bfseries C78} (2018) 203}
  [\href{https://arxiv.org/abs/1703.07364}{{\ttfamily 1703.07364}}].

\bibitem{Lin:2019uvt}
T.~Lin, \emph{{Dark matter models and direct detection}},
  \href{https://doi.org/10.22323/1.333.0009}{\emph{PoS} {\bfseries 333} (2019)
  009} [\href{https://arxiv.org/abs/1904.07915}{{\ttfamily 1904.07915}}].

\bibitem{Hooper:2018kfv}
D.~Hooper, \emph{{TASI Lectures on Indirect Searches For Dark Matter}},
  {\emph{PoS} {\bfseries TASI2018} (2019) 010}
  [\href{https://arxiv.org/abs/1812.02029}{{\ttfamily 1812.02029}}].

\bibitem{McDonald:2001vt}
J.~McDonald, \emph{{Thermally generated gauge singlet scalars as
  selfinteracting dark matter}},
  \href{https://doi.org/10.1103/PhysRevLett.88.091304}{\emph{Phys.Rev.Lett.}
  {\bfseries 88} (2002) 091304}
  [\href{https://arxiv.org/abs/hep-ph/0106249}{{\ttfamily hep-ph/0106249}}].

\bibitem{Choi:2005vq}
K.-Y. Choi and L.~Roszkowski, \emph{{E-WIMPs}},
  \href{https://doi.org/10.1063/1.2149672}{\emph{AIP Conf. Proc.} {\bfseries
  805} (2006) 30} [\href{https://arxiv.org/abs/hep-ph/0511003}{{\ttfamily
  hep-ph/0511003}}].

\bibitem{Kusenko:2006rh}
A.~Kusenko, \emph{{Sterile neutrinos, dark matter, and the pulsar velocities in
  models with a Higgs singlet}},
  \href{https://doi.org/10.1103/PhysRevLett.97.241301}{\emph{Phys. Rev. Lett.}
  {\bfseries 97} (2006) 241301}
  [\href{https://arxiv.org/abs/hep-ph/0609081}{{\ttfamily hep-ph/0609081}}].

\bibitem{Petraki:2007gq}
K.~Petraki and A.~Kusenko, \emph{{Dark-matter sterile neutrinos in models with
  a gauge singlet in the Higgs sector}},
  \href{https://doi.org/10.1103/PhysRevD.77.065014}{\emph{Phys. Rev.}
  {\bfseries D77} (2008) 065014}
  [\href{https://arxiv.org/abs/0711.4646}{{\ttfamily 0711.4646}}].

\bibitem{Hall:2009bx}
L.~J. Hall, K.~Jedamzik, J.~March-Russell and S.~M. West, \emph{{Freeze-In
  Production of FIMP Dark Matter}},
  \href{https://doi.org/10.1007/JHEP03(2010)080}{\emph{JHEP} {\bfseries 1003}
  (2010) 080} [\href{https://arxiv.org/abs/0911.1120}{{\ttfamily 0911.1120}}].

\bibitem{Chu:2011be}
X.~Chu, T.~Hambye and M.~H.~G. Tytgat, \emph{{The Four Basic Ways of Creating
  Dark Matter Through a Portal}},
  \href{https://doi.org/10.1088/1475-7516/2012/05/034}{\emph{JCAP} {\bfseries
  1205} (2012) 034} [\href{https://arxiv.org/abs/1112.0493}{{\ttfamily
  1112.0493}}].

\bibitem{Belanger:2018mqt}
G.~Bélanger, F.~Boudjema, A.~Goudelis, A.~Pukhov and B.~Zaldivar,
  \emph{{micrOMEGAs5.0 : Freeze-in}},
  \href{https://doi.org/10.1016/j.cpc.2018.04.027}{\emph{Comput. Phys. Commun.}
  {\bfseries 231} (2018) 173}
  [\href{https://arxiv.org/abs/1801.03509}{{\ttfamily 1801.03509}}].

\bibitem{Bernal:2017kxu}
N.~Bernal, M.~Heikinheimo, T.~Tenkanen, K.~Tuominen and V.~Vaskonen, \emph{{The
  Dawn of FIMP Dark Matter: A Review of Models and Constraints}},
  \href{https://doi.org/10.1142/S0217751X1730023X}{\emph{Int. J. Mod. Phys.}
  {\bfseries A32} (2017) 1730023}
  [\href{https://arxiv.org/abs/1706.07442}{{\ttfamily 1706.07442}}].

\bibitem{Davidson:2000dw}
S.~Davidson, M.~Losada and A.~Riotto, \emph{{A New perspective on
  baryogenesis}},
  \href{https://doi.org/10.1103/PhysRevLett.84.4284}{\emph{Phys. Rev. Lett.}
  {\bfseries 84} (2000) 4284}
  [\href{https://arxiv.org/abs/hep-ph/0001301}{{\ttfamily hep-ph/0001301}}].

\bibitem{Giudice:2000ex}
G.~F. Giudice, E.~W. Kolb and A.~Riotto, \emph{{Largest temperature of the
  radiation era and its cosmological implications}},
  \href{https://doi.org/10.1103/PhysRevD.64.023508}{\emph{Phys. Rev.}
  {\bfseries D64} (2001) 023508}
  [\href{https://arxiv.org/abs/hep-ph/0005123}{{\ttfamily hep-ph/0005123}}].

\bibitem{Allahverdi:2010im}
R.~Allahverdi, B.~Dutta and K.~Sinha, \emph{{Baryogenesis and Late-Decaying
  Moduli}}, \href{https://doi.org/10.1103/PhysRevD.82.035004}{\emph{Phys. Rev.}
  {\bfseries D82} (2010) 035004}
  [\href{https://arxiv.org/abs/1005.2804}{{\ttfamily 1005.2804}}].

\bibitem{Beniwal:2017eik}
A.~Beniwal, M.~Lewicki, J.~D. Wells, M.~White and A.~G. Williams,
  \emph{{Gravitational wave, collider and dark matter signals from a scalar
  singlet electroweak baryogenesis}},
  \href{https://doi.org/10.1007/JHEP08(2017)108}{\emph{JHEP} {\bfseries 08}
  (2017) 108} [\href{https://arxiv.org/abs/1702.06124}{{\ttfamily
  1702.06124}}].

\bibitem{Allahverdi:2017edd}
R.~Allahverdi, P.~S.~B. Dev and B.~Dutta, \emph{{A simple testable model of
  baryon number violation: Baryogenesis, dark matter, neutron–antineutron
  oscillation and collider signals}},
  \href{https://doi.org/10.1016/j.physletb.2018.02.019}{\emph{Phys. Lett.}
  {\bfseries B779} (2018) 262}
  [\href{https://arxiv.org/abs/1712.02713}{{\ttfamily 1712.02713}}].

\bibitem{Bernal:2017zvx}
N.~Bernal and C.~S. Fong, \emph{{Hot Leptogenesis from Thermal Dark Matter}},
  \href{https://doi.org/10.1088/1475-7516/2017/10/042}{\emph{JCAP} {\bfseries
  1710} (2017) 042} [\href{https://arxiv.org/abs/1707.02988}{{\ttfamily
  1707.02988}}].

\bibitem{Assadullahi:2009nf}
H.~Assadullahi and D.~Wands, \emph{{Gravitational waves from an early matter
  era}}, \href{https://doi.org/10.1103/PhysRevD.79.083511}{\emph{Phys. Rev.}
  {\bfseries D79} (2009) 083511}
  [\href{https://arxiv.org/abs/0901.0989}{{\ttfamily 0901.0989}}].

\bibitem{Durrer:2011bi}
R.~Durrer and J.~Hasenkamp, \emph{{Testing Superstring Theories with
  Gravitational Waves}},
  \href{https://doi.org/10.1103/PhysRevD.84.064027}{\emph{Phys. Rev.}
  {\bfseries D84} (2011) 064027}
  [\href{https://arxiv.org/abs/1105.5283}{{\ttfamily 1105.5283}}].

\bibitem{Alabidi:2013wtp}
L.~Alabidi, K.~Kohri, M.~Sasaki and Y.~Sendouda, \emph{{Observable induced
  gravitational waves from an early matter phase}},
  \href{https://doi.org/10.1088/1475-7516/2013/05/033}{\emph{JCAP} {\bfseries
  1305} (2013) 033} [\href{https://arxiv.org/abs/1303.4519}{{\ttfamily
  1303.4519}}].

\bibitem{DEramo:2019tit}
F.~D'Eramo and K.~Schmitz, \emph{{Imprint of a scalar era on the primordial
  spectrum of gravitational waves}},
  \href{https://arxiv.org/abs/1904.07870}{{\ttfamily 1904.07870}}.

\bibitem{Bernal:2019lpc}
N.~Bernal and F.~Hajkarim, \emph{{Primordial Gravitational Waves in Nonstandard
  Cosmologies}}, \href{https://doi.org/10.1103/PhysRevD.100.063502}{\emph{Phys.
  Rev.} {\bfseries D100} (2019) 063502}
  [\href{https://arxiv.org/abs/1905.10410}{{\ttfamily 1905.10410}}].

\bibitem{Figueroa:2019paj}
D.~G. Figueroa and E.~H. Tanin, \emph{{Ability of LIGO and LISA to probe the
  equation of state of the early Universe}},
  \href{https://doi.org/10.1088/1475-7516/2019/08/011}{\emph{JCAP} {\bfseries
  2019} (2020) 011} [\href{https://arxiv.org/abs/1905.11960}{{\ttfamily
  1905.11960}}].

\bibitem{Barrow:1982ei}
J.~D. Barrow, \emph{{Massive Particles as a Probe of the Early Universe}},
  \href{https://doi.org/10.1016/0550-3213(82)90233-4}{\emph{Nucl. Phys.}
  {\bfseries B208} (1982) 501}.

\bibitem{Ford:1986sy}
L.~H. Ford, \emph{{Gravitational Particle Creation and Inflation}},
  \href{https://doi.org/10.1103/PhysRevD.35.2955}{\emph{Phys. Rev.} {\bfseries
  D35} (1987) 2955}.

\bibitem{Kane:2015jia}
G.~Kane, K.~Sinha and S.~Watson, \emph{{Cosmological Moduli and the
  Post-Inflationary Universe: A Critical Review}},
  \href{https://doi.org/10.1142/S0218271815300220}{\emph{Int. J. Mod. Phys.}
  {\bfseries D24} (2015) 1530022}
  [\href{https://arxiv.org/abs/1502.07746}{{\ttfamily 1502.07746}}].

\bibitem{Co:2015pka}
R.~T. Co, F.~D'Eramo, L.~J. Hall and D.~Pappadopulo, \emph{{Freeze-In Dark
  Matter with Displaced Signatures at Colliders}},
  \href{https://doi.org/10.1088/1475-7516/2015/12/024}{\emph{JCAP} {\bfseries
  1512} (2015) 024} [\href{https://arxiv.org/abs/1506.07532}{{\ttfamily
  1506.07532}}].

\bibitem{Davoudiasl:2015vba}
H.~Davoudiasl, D.~Hooper and S.~D. McDermott, \emph{{Inflatable Dark Matter}},
  \href{https://doi.org/10.1103/PhysRevLett.116.031303}{\emph{Phys. Rev. Lett.}
  {\bfseries 116} (2016) 031303}
  [\href{https://arxiv.org/abs/1507.08660}{{\ttfamily 1507.08660}}].

\bibitem{Randall:2015xza}
L.~Randall, J.~Scholtz and J.~Unwin, \emph{{Flooded Dark Matter and S Level
  Rise}}, \href{https://doi.org/10.1007/JHEP03(2016)011}{\emph{JHEP} {\bfseries
  03} (2016) 011} [\href{https://arxiv.org/abs/1509.08477}{{\ttfamily
  1509.08477}}].

\bibitem{Berlin:2016vnh}
A.~Berlin, D.~Hooper and G.~Krnjaic, \emph{{PeV-Scale Dark Matter as a Thermal
  Relic of a Decoupled Sector}},
  \href{https://doi.org/10.1016/j.physletb.2016.06.037}{\emph{Phys. Lett.}
  {\bfseries B760} (2016) 106}
  [\href{https://arxiv.org/abs/1602.08490}{{\ttfamily 1602.08490}}].

\bibitem{Tenkanen:2016jic}
T.~Tenkanen and V.~Vaskonen, \emph{{Reheating the Standard Model from a hidden
  sector}}, \href{https://doi.org/10.1103/PhysRevD.94.083516}{\emph{Phys. Rev.}
  {\bfseries D94} (2016) 083516}
  [\href{https://arxiv.org/abs/1606.00192}{{\ttfamily 1606.00192}}].

\bibitem{Dror:2016rxc}
J.~A. Dror, E.~Kuflik and W.~H. Ng, \emph{{Codecaying Dark Matter}},
  \href{https://doi.org/10.1103/PhysRevLett.117.211801}{\emph{Phys. Rev. Lett.}
  {\bfseries 117} (2016) 211801}
  [\href{https://arxiv.org/abs/1607.03110}{{\ttfamily 1607.03110}}].

\bibitem{Berlin:2016gtr}
A.~Berlin, D.~Hooper and G.~Krnjaic, \emph{{Thermal Dark Matter From A Highly
  Decoupled Sector}},
  \href{https://doi.org/10.1103/PhysRevD.94.095019}{\emph{Phys. Rev.}
  {\bfseries D94} (2016) 095019}
  [\href{https://arxiv.org/abs/1609.02555}{{\ttfamily 1609.02555}}].

\bibitem{DEramo:2017gpl}
F.~D'Eramo, N.~Fernandez and S.~Profumo, \emph{{When the Universe Expands Too
  Fast: Relentless Dark Matter}},
  \href{https://doi.org/10.1088/1475-7516/2017/05/012}{\emph{JCAP} {\bfseries
  1705} (2017) 012} [\href{https://arxiv.org/abs/1703.04793}{{\ttfamily
  1703.04793}}].

\bibitem{Hamdan:2017psw}
S.~Hamdan and J.~Unwin, \emph{{Dark Matter Freeze-out During Matter
  Domination}}, \href{https://doi.org/10.1142/S021773231850181X}{\emph{Mod.
  Phys. Lett.} {\bfseries A33} (2018) 1850181}
  [\href{https://arxiv.org/abs/1710.03758}{{\ttfamily 1710.03758}}].

\bibitem{Visinelli:2017qga}
L.~Visinelli, \emph{{(Non-)thermal production of WIMPs during kination}},
  \href{https://doi.org/10.3390/sym10110546}{\emph{Symmetry} {\bfseries 10}
  (2018) 546} [\href{https://arxiv.org/abs/1710.11006}{{\ttfamily
  1710.11006}}].

\bibitem{Dror:2017gjq}
J.~A. Dror, E.~Kuflik, B.~Melcher and S.~Watson, \emph{{Concentrated Dark
  Matter: Enhanced Small-scale Structure from Co-Decaying Dark Matter}},
  \href{https://doi.org/10.1103/PhysRevD.97.063524}{\emph{Phys. Rev.}
  {\bfseries D97} (2018) 063524}
  [\href{https://arxiv.org/abs/1711.04773}{{\ttfamily 1711.04773}}].

\bibitem{Drees:2017iod}
M.~Drees and F.~Hajkarim, \emph{{Dark Matter Production in an Early Matter
  Dominated Era}},
  \href{https://doi.org/10.1088/1475-7516/2018/02/057}{\emph{JCAP} {\bfseries
  1802} (2018) 057} [\href{https://arxiv.org/abs/1711.05007}{{\ttfamily
  1711.05007}}].

\bibitem{DEramo:2017ecx}
F.~D'Eramo, N.~Fernandez and S.~Profumo, \emph{{Dark Matter Freeze-in
  Production in Fast-Expanding Universes}},
  \href{https://doi.org/10.1088/1475-7516/2018/02/046}{\emph{JCAP} {\bfseries
  1802} (2018) 046} [\href{https://arxiv.org/abs/1712.07453}{{\ttfamily
  1712.07453}}].

\bibitem{Maity:2018dgy}
D.~Maity and P.~Saha, \emph{{Connecting CMB anisotropy and cold dark matter
  phenomenology via reheating}},
  \href{https://doi.org/10.1103/PhysRevD.98.103525}{\emph{Phys. Rev.}
  {\bfseries D98} (2018) 103525}
  [\href{https://arxiv.org/abs/1801.03059}{{\ttfamily 1801.03059}}].

\bibitem{Bernal:2018ins}
N.~Bernal, C.~Cosme and T.~Tenkanen, \emph{{Phenomenology of Self-Interacting
  Dark Matter in a Matter-Dominated Universe}},
  \href{https://doi.org/10.1140/epjc/s10052-019-6608-8}{\emph{Eur. Phys. J.}
  {\bfseries C79} (2019) 99}
  [\href{https://arxiv.org/abs/1803.08064}{{\ttfamily 1803.08064}}].

\bibitem{Hardy:2018bph}
E.~Hardy, \emph{{Higgs portal dark matter in non-standard cosmological
  histories}}, \href{https://doi.org/10.1007/JHEP06(2018)043}{\emph{JHEP}
  {\bfseries 06} (2018) 043}
  [\href{https://arxiv.org/abs/1804.06783}{{\ttfamily 1804.06783}}].

\bibitem{Maity:2018exj}
D.~Maity and P.~Saha, \emph{{CMB constraints on dark matter phenomenology via
  reheating in Minimal plateau inflation}},
  \href{https://doi.org/10.1016/j.dark.2019.100317}{\emph{Phys. Dark Univ.}
  (2018) 100317} [\href{https://arxiv.org/abs/1804.10115}{{\ttfamily
  1804.10115}}].

\bibitem{Hambye:2018qjv}
T.~Hambye, A.~Strumia and D.~Teresi, \emph{{Super-cool Dark Matter}},
  \href{https://doi.org/10.1007/JHEP08(2018)188}{\emph{JHEP} {\bfseries 08}
  (2018) 188} [\href{https://arxiv.org/abs/1805.01473}{{\ttfamily
  1805.01473}}].

\bibitem{Bernal:2018kcw}
N.~Bernal, C.~Cosme, T.~Tenkanen and V.~Vaskonen, \emph{{Scalar singlet dark
  matter in non-standard cosmologies}},
  \href{https://doi.org/10.1140/epjc/s10052-019-6550-9}{\emph{Eur. Phys. J.}
  {\bfseries C79} (2019) 30}
  [\href{https://arxiv.org/abs/1806.11122}{{\ttfamily 1806.11122}}].

\bibitem{Arbey:2018uho}
A.~Arbey, J.~Ellis, F.~Mahmoudi and G.~Robbins, \emph{{Dark Matter Casts Light
  on the Early Universe}},
  \href{https://doi.org/10.1007/JHEP10(2018)132}{\emph{JHEP} {\bfseries 10}
  (2018) 132} [\href{https://arxiv.org/abs/1807.00554}{{\ttfamily
  1807.00554}}].

\bibitem{Drees:2018dsj}
M.~Drees and F.~Hajkarim, \emph{{Neutralino Dark Matter in Scenarios with Early
  Matter Domination}},
  \href{https://doi.org/10.1007/JHEP12(2018)042}{\emph{JHEP} {\bfseries 12}
  (2018) 042} [\href{https://arxiv.org/abs/1808.05706}{{\ttfamily
  1808.05706}}].

\bibitem{Betancur:2018xtj}
A.~Betancur and {\'O}.~Zapata, \emph{{Phenomenology of doublet-triplet
  fermionic dark matter in nonstandard cosmology and multicomponent dark
  sectors}}, \href{https://doi.org/10.1103/PhysRevD.98.095003}{\emph{Phys.
  Rev.} {\bfseries D98} (2018) 095003}
  [\href{https://arxiv.org/abs/1809.04990}{{\ttfamily 1809.04990}}].

\bibitem{Maldonado:2019qmp}
C.~Maldonado and J.~Unwin, \emph{{Establishing the Dark Matter Relic Density in
  an Era of Particle Decays}},
  \href{https://doi.org/10.1088/1475-7516/2019/06/037}{\emph{JCAP} {\bfseries
  1906} (2019) 037} [\href{https://arxiv.org/abs/1902.10746}{{\ttfamily
  1902.10746}}].

\bibitem{Poulin:2019omz}
A.~Poulin, \emph{{Dark matter freeze-out in modified cosmological scenarios}},
  \href{https://doi.org/10.1103/PhysRevD.100.043022}{\emph{Phys. Rev.}
  {\bfseries D100} (2019) 043022}
  [\href{https://arxiv.org/abs/1905.03126}{{\ttfamily 1905.03126}}].

\bibitem{Tenkanen:2019cik}
T.~Tenkanen, \emph{{The Standard Model Higgs and Hidden Sector Cosmology}},
  \href{https://arxiv.org/abs/1905.11737}{{\ttfamily 1905.11737}}.

\bibitem{Kamionkowski:1990ni}
M.~Kamionkowski and M.~S. Turner, \emph{{Thermal Relics: Do we Know their
  Abundances?}}, \href{https://doi.org/10.1103/PhysRevD.42.3310}{\emph{Phys.
  Rev.} {\bfseries D42} (1990) 3310}.

\bibitem{McDonald:1989jd}
J.~McDonald, \emph{{{WIMP} Densities in Decaying Particle Dominated
  Cosmology}}, \href{https://doi.org/10.1103/PhysRevD.43.1063}{\emph{Phys.
  Rev.} {\bfseries D43} (1991) 1063}.

\bibitem{Salati:2002md}
P.~Salati, \emph{{Quintessence and the relic density of neutralinos}},
  \href{https://doi.org/10.1016/j.physletb.2003.07.073}{\emph{Phys. Lett.}
  {\bfseries B571} (2003) 121}
  [\href{https://arxiv.org/abs/astro-ph/0207396}{{\ttfamily
  astro-ph/0207396}}].

\bibitem{Comelli:2003cv}
D.~Comelli, M.~Pietroni and A.~Riotto, \emph{{Dark energy and dark matter}},
  \href{https://doi.org/10.1016/j.physletb.2003.05.006}{\emph{Phys. Lett.}
  {\bfseries B571} (2003) 115}
  [\href{https://arxiv.org/abs/hep-ph/0302080}{{\ttfamily hep-ph/0302080}}].

\bibitem{Rosati:2003yw}
F.~Rosati, \emph{{Quintessential enhancement of dark matter abundance}},
  \href{https://doi.org/10.1016/j.physletb.2003.07.048}{\emph{Phys. Lett.}
  {\bfseries B570} (2003) 5}
  [\href{https://arxiv.org/abs/hep-ph/0302159}{{\ttfamily hep-ph/0302159}}].

\bibitem{Pallis:2004yy}
C.~Pallis, \emph{{Massive particle decay and cold dark matter abundance}},
  \href{https://doi.org/10.1016/j.astropartphys.2004.05.006}{\emph{Astropart.
  Phys.} {\bfseries 21} (2004) 689}
  [\href{https://arxiv.org/abs/hep-ph/0402033}{{\ttfamily hep-ph/0402033}}].

\bibitem{Gelmini:2006pw}
G.~B. Gelmini and P.~Gondolo, \emph{{Neutralino with the right cold dark matter
  abundance in (almost) any supersymmetric model}},
  \href{https://doi.org/10.1103/PhysRevD.74.023510}{\emph{Phys. Rev.}
  {\bfseries D74} (2006) 023510}
  [\href{https://arxiv.org/abs/hep-ph/0602230}{{\ttfamily hep-ph/0602230}}].

\bibitem{Gelmini:2006pq}
G.~Gelmini, P.~Gondolo, A.~Soldatenko and C.~E. Yaguna, \emph{{The Effect of a
  late decaying scalar on the neutralino relic density}},
  \href{https://doi.org/10.1103/PhysRevD.74.083514}{\emph{Phys. Rev.}
  {\bfseries D74} (2006) 083514}
  [\href{https://arxiv.org/abs/hep-ph/0605016}{{\ttfamily hep-ph/0605016}}].

\bibitem{Arbey:2008kv}
A.~Arbey and F.~Mahmoudi, \emph{{SUSY constraints from relic density: High
  sensitivity to pre-BBN expansion rate}},
  \href{https://doi.org/10.1016/j.physletb.2008.09.032}{\emph{Phys. Lett.}
  {\bfseries B669} (2008) 46}
  [\href{https://arxiv.org/abs/0803.0741}{{\ttfamily 0803.0741}}].

\bibitem{Cohen:2008nb}
T.~Cohen, D.~E. Morrissey and A.~Pierce, \emph{{Changes in Dark Matter
  Properties After Freeze-Out}},
  \href{https://doi.org/10.1103/PhysRevD.78.111701}{\emph{Phys. Rev.}
  {\bfseries D78} (2008) 111701}
  [\href{https://arxiv.org/abs/0808.3994}{{\ttfamily 0808.3994}}].

\bibitem{Arbey:2009gt}
A.~Arbey and F.~Mahmoudi, \emph{{SUSY Constraints, Relic Density, and Very
  Early Universe}}, \href{https://doi.org/10.1007/JHEP05(2010)051}{\emph{JHEP}
  {\bfseries 05} (2010) 051} [\href{https://arxiv.org/abs/0906.0368}{{\ttfamily
  0906.0368}}].

\bibitem{Chung:1998rq}
D.~J.~H. Chung, E.~W. Kolb and A.~Riotto, \emph{{Production of massive
  particles during reheating}},
  \href{https://doi.org/10.1103/PhysRevD.60.063504}{\emph{Phys. Rev.}
  {\bfseries D60} (1999) 063504}
  [\href{https://arxiv.org/abs/hep-ph/9809453}{{\ttfamily hep-ph/9809453}}].

\bibitem{Kolb:2003ke}
E.~W. Kolb, A.~Notari and A.~Riotto, \emph{{On the reheating stage after
  inflation}}, \href{https://doi.org/10.1103/PhysRevD.68.123505}{\emph{Phys.
  Rev.} {\bfseries D68} (2003) 123505}
  [\href{https://arxiv.org/abs/hep-ph/0307241}{{\ttfamily hep-ph/0307241}}].

\bibitem{Garcia:2017tuj}
M.~A.~G. Garcia, Y.~Mambrini, K.~A. Olive and M.~Peloso, \emph{{Enhancement of
  the Dark Matter Abundance Before Reheating: Applications to Gravitino Dark
  Matter}}, \href{https://doi.org/10.1103/PhysRevD.96.103510}{\emph{Phys. Rev.}
  {\bfseries D96} (2017) 103510}
  [\href{https://arxiv.org/abs/1709.01549}{{\ttfamily 1709.01549}}].

\bibitem{Ellis:2015jpg}
J.~Ellis, M.~A.~G. Garcia, D.~V. Nanopoulos, K.~A. Olive and M.~Peloso,
  \emph{{Post-Inflationary Gravitino Production Revisited}},
  \href{https://doi.org/10.1088/1475-7516/2016/03/008}{\emph{JCAP} {\bfseries
  1603} (2016) 008} [\href{https://arxiv.org/abs/1512.05701}{{\ttfamily
  1512.05701}}].

\bibitem{Gelmini:2008sh}
G.~B. Gelmini and P.~Gondolo, \emph{{Ultra-cold WIMPs: relics of non-standard
  pre-BBN cosmologies}},
  \href{https://doi.org/10.1088/1475-7516/2008/10/002}{\emph{JCAP} {\bfseries
  0810} (2008) 002} [\href{https://arxiv.org/abs/0803.2349}{{\ttfamily
  0803.2349}}].

\bibitem{Visinelli:2015eka}
L.~Visinelli and P.~Gondolo, \emph{{Kinetic decoupling of WIMPs: analytic
  expressions}}, \href{https://doi.org/10.1103/PhysRevD.91.083526}{\emph{Phys.
  Rev.} {\bfseries D91} (2015) 083526}
  [\href{https://arxiv.org/abs/1501.02233}{{\ttfamily 1501.02233}}].

\bibitem{Waldstein:2016blt}
I.~R. Waldstein, A.~L. Erickcek and C.~Ilie, \emph{{Quasidecoupled state for
  dark matter in nonstandard thermal histories}},
  \href{https://doi.org/10.1103/PhysRevD.95.123531}{\emph{Phys. Rev.}
  {\bfseries D95} (2017) 123531}
  [\href{https://arxiv.org/abs/1609.05927}{{\ttfamily 1609.05927}}].

\bibitem{Waldstein:2017wps}
I.~R. Waldstein and A.~L. Erickcek, \emph{{Comment on “Kinetic decoupling of
  WIMPs: Analytic expressions”}},
  \href{https://doi.org/10.1103/PhysRevD.95.088301}{\emph{Phys. Rev.}
  {\bfseries D95} (2017) 088301}
  [\href{https://arxiv.org/abs/1707.03417}{{\ttfamily 1707.03417}}].

\bibitem{Steigman:2012nb}
G.~Steigman, B.~Dasgupta and J.~F. Beacom, \emph{{Precise Relic WIMP Abundance
  and its Impact on Searches for Dark Matter Annihilation}},
  \href{https://doi.org/10.1103/PhysRevD.86.023506}{\emph{Phys. Rev.}
  {\bfseries D86} (2012) 023506}
  [\href{https://arxiv.org/abs/1204.3622}{{\ttfamily 1204.3622}}].

\bibitem{Mena:2007ty}
O.~Mena, S.~Palomares-Ruiz and S.~Pascoli, \emph{{Reconstructing WIMP
  properties with neutrino detectors}},
  \href{https://doi.org/10.1016/j.physletb.2008.04.059}{\emph{Phys. Lett.}
  {\bfseries B664} (2008) 92}
  [\href{https://arxiv.org/abs/0706.3909}{{\ttfamily 0706.3909}}].

\bibitem{Drees:2008bv}
M.~Drees and C.-L. Shan, \emph{{Model-Independent Determination of the WIMP
  Mass from Direct Dark Matter Detection Data}},
  \href{https://doi.org/10.1088/1475-7516/2008/06/012}{\emph{JCAP} {\bfseries
  0806} (2008) 012} [\href{https://arxiv.org/abs/0803.4477}{{\ttfamily
  0803.4477}}].

\bibitem{Bernal:2008zk}
N.~Bernal, A.~Goudelis, Y.~Mambrini and C.~Muñoz, \emph{{Determining the WIMP
  mass using the complementarity between direct and indirect searches and the
  ILC}}, \href{https://doi.org/10.1088/1475-7516/2009/01/046}{\emph{JCAP}
  {\bfseries 0901} (2009) 046}
  [\href{https://arxiv.org/abs/0804.1976}{{\ttfamily 0804.1976}}].

\bibitem{Bernal:2008cu}
N.~Bernal, \emph{{WIMP mass from direct, indirect dark matter detection
  experiments and colliders: A Complementary and model-independent approach}},
  in \emph{{Proceedings, $43^\text{rd}$ Rencontres de Moriond on Electroweak
  Interactions and Unified Theories: La Thuile, Italy, March 1-8, 2008}}, 2008,
  \href{https://arxiv.org/abs/0805.2241}{{\ttfamily 0805.2241}},
  \href{https://inspirehep.net/record/785807/files/arXiv:0805.2241.pdf}{https://inspirehep.net/record/785807/files/arXiv:0805.2241.pdf}.

\bibitem{Bergstrom:2010gh}
L.~Bergström, T.~Bringmann and J.~Edsjö, \emph{{Complementarity of direct
  dark matter detection and indirect detection through gamma-rays}},
  \href{https://doi.org/10.1103/PhysRevD.83.045024}{\emph{Phys. Rev.}
  {\bfseries D83} (2011) 045024}
  [\href{https://arxiv.org/abs/1011.4514}{{\ttfamily 1011.4514}}].

\bibitem{Pato:2010zk}
M.~Pato, L.~Baudis, G.~Bertone, R.~Ruiz~de Austri, L.~E. Strigari and
  R.~Trotta, \emph{{Complementarity of Dark Matter Direct Detection Targets}},
  \href{https://doi.org/10.1103/PhysRevD.83.083505}{\emph{Phys. Rev.}
  {\bfseries D83} (2011) 083505}
  [\href{https://arxiv.org/abs/1012.3458}{{\ttfamily 1012.3458}}].

\bibitem{Arisaka:2011eu}
K.~Arisaka et~al., \emph{{Studies of a three-stage dark matter and neutrino
  observatory based on multi-ton combinations of liquid xenon and liquid argon
  detectors}},
  \href{https://doi.org/10.1016/j.astropartphys.2012.05.006}{\emph{Astropart.
  Phys.} {\bfseries 36} (2012) 93}
  [\href{https://arxiv.org/abs/1107.1295}{{\ttfamily 1107.1295}}].

\bibitem{Cerdeno:2013gqa}
D.~G. Cerdeño et~al., \emph{{Complementarity of dark matter direct detection:
  the role of bolometric targets}},
  \href{https://doi.org/10.1088/1475-7516/2013/07/028,
  10.1088/1475-7516/2013/09/E01}{\emph{JCAP} {\bfseries 1307} (2013) 028}
  [\href{https://arxiv.org/abs/1304.1758}{{\ttfamily 1304.1758}}].

\bibitem{Arina:2013jya}
C.~Arina, G.~Bertone and H.~Silverwood, \emph{{Complementarity of direct and
  indirect Dark Matter detection experiments}},
  \href{https://doi.org/10.1103/PhysRevD.88.013002}{\emph{Phys. Rev.}
  {\bfseries D88} (2013) 013002}
  [\href{https://arxiv.org/abs/1304.5119}{{\ttfamily 1304.5119}}].

\bibitem{Peter:2013aha}
A.~H.~G. Peter, V.~Gluscevic, A.~M. Green, B.~J. Kavanagh and S.~K. Lee,
  \emph{{WIMP physics with ensembles of direct-detection experiments}},
  \href{https://doi.org/10.1016/j.dark.2014.10.006}{\emph{Phys. Dark Univ.}
  {\bfseries 5-6} (2014) 45} [\href{https://arxiv.org/abs/1310.7039}{{\ttfamily
  1310.7039}}].

\bibitem{Kavanagh:2014rya}
B.~J. Kavanagh, M.~Fornasa and A.~M. Green, \emph{{Probing WIMP particle
  physics and astrophysics with direct detection and neutrino telescope data}},
  \href{https://doi.org/10.1103/PhysRevD.91.103533}{\emph{Phys. Rev.}
  {\bfseries D91} (2015) 103533}
  [\href{https://arxiv.org/abs/1410.8051}{{\ttfamily 1410.8051}}].

\bibitem{Roszkowski:2016bhs}
L.~Roszkowski, E.~M. Sessolo, S.~Trojanowski and A.~J. Williams,
  \emph{{Reconstructing WIMP properties through an interplay of signal
  measurements in direct detection, Fermi-LAT, and CTA searches for dark
  matter}}, \href{https://doi.org/10.1088/1475-7516/2016/08/033}{\emph{JCAP}
  {\bfseries 1608} (2016) 033}
  [\href{https://arxiv.org/abs/1603.06519}{{\ttfamily 1603.06519}}].

\bibitem{Queiroz:2016sxf}
F.~S. Queiroz, W.~Rodejohann and C.~E. Yaguna, \emph{{Is the dark matter
  particle its own antiparticle?}},
  \href{https://doi.org/10.1103/PhysRevD.95.095010}{\emph{Phys. Rev.}
  {\bfseries D95} (2017) 095010}
  [\href{https://arxiv.org/abs/1610.06581}{{\ttfamily 1610.06581}}].

\bibitem{Roszkowski:2017dou}
L.~Roszkowski, S.~Trojanowski and K.~Turzyński, \emph{{Towards understanding
  thermal history of the Universe through direct and indirect detection of dark
  matter}}, \href{https://doi.org/10.1088/1475-7516/2017/10/005}{\emph{JCAP}
  {\bfseries 1710} (2017) 005}
  [\href{https://arxiv.org/abs/1703.00841}{{\ttfamily 1703.00841}}].

\bibitem{Kavanagh:2017hcl}
B.~J. Kavanagh, F.~S. Queiroz, W.~Rodejohann and C.~E. Yaguna, \emph{{Prospects
  for determining the particle/antiparticle nature of WIMP dark matter with
  direct detection experiments}},
  \href{https://doi.org/10.1007/JHEP10(2017)059}{\emph{JHEP} {\bfseries 10}
  (2017) 059} [\href{https://arxiv.org/abs/1706.07819}{{\ttfamily
  1706.07819}}].

\bibitem{Bertone:2017adx}
G.~Bertone, N.~Bozorgnia, J.~S. Kim, S.~Liem, C.~McCabe, S.~Otten et~al.,
  \emph{{Identifying WIMP dark matter from particle and astroparticle data}},
  \href{https://doi.org/10.1088/1475-7516/2018/03/026}{\emph{JCAP} {\bfseries
  1803} (2018) 026} [\href{https://arxiv.org/abs/1712.04793}{{\ttfamily
  1712.04793}}].

\bibitem{Queiroz:2018utk}
F.~S. Queiroz and C.~E. Yaguna, \emph{{Gamma-ray lines may reveal the CP nature
  of the dark matter particle}},
  \href{https://doi.org/10.1088/1475-7516/2019/01/047}{\emph{JCAP} {\bfseries
  1901} (2019) 047} [\href{https://arxiv.org/abs/1810.07068}{{\ttfamily
  1810.07068}}].

\bibitem{Green:2002ht}
A.~M. Green, \emph{{Effect of halo modeling on WIMP exclusion limits}},
  \href{https://doi.org/10.1103/PhysRevD.66.083003}{\emph{Phys. Rev.}
  {\bfseries D66} (2002) 083003}
  [\href{https://arxiv.org/abs/astro-ph/0207366}{{\ttfamily
  astro-ph/0207366}}].

\bibitem{Zemp:2008gw}
M.~Zemp, J.~Diemand, M.~Kuhlen, P.~Madau, B.~Moore, D.~Potter et~al.,
  \emph{{The Graininess of Dark Matter Haloes}},
  \href{https://doi.org/10.1111/j.1365-2966.2008.14361.x}{\emph{Mon. Not. Roy.
  Astron. Soc.} {\bfseries 394} (2009) 641}
  [\href{https://arxiv.org/abs/0812.2033}{{\ttfamily 0812.2033}}].

\bibitem{McCabe:2010zh}
C.~McCabe, \emph{{The Astrophysical Uncertainties Of Dark Matter Direct
  Detection Experiments}},
  \href{https://doi.org/10.1103/PhysRevD.82.023530}{\emph{Phys. Rev.}
  {\bfseries D82} (2010) 023530}
  [\href{https://arxiv.org/abs/1005.0579}{{\ttfamily 1005.0579}}].

\bibitem{Bernal:2010ip}
N.~Bernal and S.~Palomares-Ruiz, \emph{{Constraining Dark Matter Properties
  with Gamma-Rays from the Galactic Center with Fermi-LAT}},
  \href{https://doi.org/10.1016/j.nuclphysb.2011.12.016}{\emph{Nucl. Phys.}
  {\bfseries B857} (2012) 380}
  [\href{https://arxiv.org/abs/1006.0477}{{\ttfamily 1006.0477}}].

\bibitem{Pato:2010yq}
M.~Pato, O.~Agertz, G.~Bertone, B.~Moore and R.~Teyssier, \emph{{Systematic
  uncertainties in the determination of the local dark matter density}},
  \href{https://doi.org/10.1103/PhysRevD.82.023531}{\emph{Phys. Rev.}
  {\bfseries D82} (2010) 023531}
  [\href{https://arxiv.org/abs/1006.1322}{{\ttfamily 1006.1322}}].

\bibitem{Bernal:2011pz}
N.~Bernal and S.~Palomares-Ruiz, \emph{{Constraining the Milky Way Dark Matter
  Density Profile with Gamma-Rays with Fermi-LAT}},
  \href{https://doi.org/10.1088/1475-7516/2012/01/006}{\emph{JCAP} {\bfseries
  1201} (2012) 006} [\href{https://arxiv.org/abs/1103.2377}{{\ttfamily
  1103.2377}}].

\bibitem{Fairbairn:2012zs}
M.~Fairbairn, T.~Douce and J.~Swift, \emph{{Quantifying Astrophysical
  Uncertainties on Dark Matter Direct Detection Results}},
  \href{https://doi.org/10.1016/j.astropartphys.2013.06.003}{\emph{Astropart.
  Phys.} {\bfseries 47} (2013) 45}
  [\href{https://arxiv.org/abs/1206.2693}{{\ttfamily 1206.2693}}].

\bibitem{Bernal:2014mmt}
N.~Bernal, J.~E. Forero-Romero, R.~Garani and S.~Palomares-Ruiz,
  \emph{{Systematic uncertainties from halo asphericity in dark matter
  searches}}, \href{https://doi.org/10.1088/1475-7516/2014/09/004}{\emph{JCAP}
  {\bfseries 1409} (2014) 004}
  [\href{https://arxiv.org/abs/1405.6240}{{\ttfamily 1405.6240}}].

\bibitem{Bernal:2015oyn}
N.~Bernal, J.~E. Forero-Romero, R.~Garani and S.~Palomares-Ruiz,
  \emph{{Systematic uncertainties from halo asphericity in dark matter
  searches}},
  \href{https://doi.org/10.1016/j.nuclphysbps.2015.10.129}{\emph{Nucl. Part.
  Phys. Proc.} {\bfseries 267-269} (2015) 345}.

\bibitem{Bernal:2016guq}
N.~Bernal, L.~Necib and T.~R. Slatyer, \emph{{Spherical Cows in Dark Matter
  Indirect Detection}},
  \href{https://doi.org/10.1088/1475-7516/2016/12/030}{\emph{JCAP} {\bfseries
  1612} (2016) 030} [\href{https://arxiv.org/abs/1606.00433}{{\ttfamily
  1606.00433}}].

\bibitem{Benito:2016kyp}
M.~Benito, N.~Bernal, N.~Bozorgnia, F.~Calore and F.~Iocco, \emph{{Particle
  Dark Matter Constraints: the Effect of Galactic Uncertainties}},
  \href{https://doi.org/10.1088/1475-7516/2017/02/007}{\emph{JCAP} {\bfseries
  1702} (2017) 007} [\href{https://arxiv.org/abs/1612.02010}{{\ttfamily
  1612.02010}}].

\bibitem{Green:2017odb}
A.~M. Green, \emph{{Astrophysical uncertainties on the local dark matter
  distribution and direct detection experiments}},
  \href{https://doi.org/10.1088/1361-6471/aa7819}{\emph{J. Phys.} {\bfseries
  G44} (2017) 084001} [\href{https://arxiv.org/abs/1703.10102}{{\ttfamily
  1703.10102}}].

\bibitem{Ibarra:2018yxq}
A.~Ibarra, B.~J. Kavanagh and A.~Rappelt, \emph{{Bracketing the impact of
  astrophysical uncertainties on local dark matter searches}},
  \href{https://doi.org/10.1088/1475-7516/2018/12/018}{\emph{JCAP} {\bfseries
  1812} (2018) 018} [\href{https://arxiv.org/abs/1806.08714}{{\ttfamily
  1806.08714}}].

\bibitem{Benito:2019ngh}
M.~Benito, A.~Cuoco and F.~Iocco, \emph{{Handling the Uncertainties in the
  Galactic Dark Matter Distribution for Particle Dark Matter Searches}},
  \href{https://doi.org/10.1088/1475-7516/2019/03/033}{\emph{JCAP} {\bfseries
  1903} (2019) 033} [\href{https://arxiv.org/abs/1901.02460}{{\ttfamily
  1901.02460}}].

\bibitem{Karukes:2019jxv}
E.~V. Karukes, M.~Benito, F.~Iocco, R.~Trotta and A.~Geringer-Sameth,
  \emph{{Bayesian reconstruction of the Milky Way dark matter distribution}},
  \href{https://arxiv.org/abs/1901.02463}{{\ttfamily 1901.02463}}.

\bibitem{Wu:2019nhd}
Y.~Wu, K.~Freese, C.~Kelso and P.~Stengel, \emph{{Uncertainties in Direct Dark
  Matter Detection in Light of GAIA}},
  \href{https://arxiv.org/abs/1904.04781}{{\ttfamily 1904.04781}}.

\bibitem{Kane:2015qea}
G.~L. Kane, P.~Kumar, B.~D. Nelson and B.~Zheng, \emph{{Dark matter production
  mechanisms with a nonthermal cosmological history: A classification}},
  \href{https://doi.org/10.1103/PhysRevD.93.063527}{\emph{Phys. Rev.}
  {\bfseries D93} (2016) 063527}
  [\href{https://arxiv.org/abs/1502.05406}{{\ttfamily 1502.05406}}].

\bibitem{Drees:2015exa}
M.~Drees, F.~Hajkarim and E.~R. Schmitz, \emph{{The Effects of QCD Equation of
  State on the Relic Density of WIMP Dark Matter}},
  \href{https://doi.org/10.1088/1475-7516/2015/06/025}{\emph{JCAP} {\bfseries
  1506} (2015) 025} [\href{https://arxiv.org/abs/1503.03513}{{\ttfamily
  1503.03513}}].

\bibitem{Kawasaki:2000en}
M.~Kawasaki, K.~Kohri and N.~Sugiyama, \emph{{MeV scale reheating temperature
  and thermalization of neutrino background}},
  \href{https://doi.org/10.1103/PhysRevD.62.023506}{\emph{Phys. Rev.}
  {\bfseries D62} (2000) 023506}
  [\href{https://arxiv.org/abs/astro-ph/0002127}{{\ttfamily
  astro-ph/0002127}}].

\bibitem{Hannestad:2004px}
S.~Hannestad, \emph{{What is the lowest possible reheating temperature?}},
  \href{https://doi.org/10.1103/PhysRevD.70.043506}{\emph{Phys. Rev.}
  {\bfseries D70} (2004) 043506}
  [\href{https://arxiv.org/abs/astro-ph/0403291}{{\ttfamily
  astro-ph/0403291}}].

\bibitem{Ichikawa:2005vw}
K.~Ichikawa, M.~Kawasaki and F.~Takahashi, \emph{{The Oscillation effects on
  thermalization of the neutrinos in the Universe with low reheating
  temperature}}, \href{https://doi.org/10.1103/PhysRevD.72.043522}{\emph{Phys.
  Rev.} {\bfseries D72} (2005) 043522}
  [\href{https://arxiv.org/abs/astro-ph/0505395}{{\ttfamily
  astro-ph/0505395}}].

\bibitem{DeBernardis:2008zz}
F.~De~Bernardis, L.~Pagano and A.~Melchiorri, \emph{{New constraints on the
  reheating temperature of the universe after WMAP-5}},
  \href{https://doi.org/10.1016/j.astropartphys.2008.09.005}{\emph{Astropart.
  Phys.} {\bfseries 30} (2008) 192}.

\bibitem{Slatyer:2015jla}
T.~R. Slatyer, \emph{{Indirect dark matter signatures in the cosmic dark ages.
  I. Generalizing the bound on s-wave dark matter annihilation from Planck
  results}}, \href{https://doi.org/10.1103/PhysRevD.93.023527}{\emph{Phys.
  Rev.} {\bfseries D93} (2016) 023527}
  [\href{https://arxiv.org/abs/1506.03811}{{\ttfamily 1506.03811}}].

\bibitem{TheFermi-LAT:2017vmf}
{\scshape Fermi-LAT} collaboration, \emph{{The Fermi Galactic Center GeV Excess
  and Implications for Dark Matter}},
  \href{https://doi.org/10.3847/1538-4357/aa6cab}{\emph{Astrophys. J.}
  {\bfseries 840} (2017) 43}
  [\href{https://arxiv.org/abs/1704.03910}{{\ttfamily 1704.03910}}].

\bibitem{Fermi-LAT:2016uux}
{\scshape DES, Fermi-LAT} collaboration, \emph{{Searching for Dark Matter
  Annihilation in Recently Discovered Milky Way Satellites with Fermi-LAT}},
  \href{https://doi.org/10.3847/1538-4357/834/2/110}{\emph{Astrophys. J.}
  {\bfseries 834} (2017) 110}
  [\href{https://arxiv.org/abs/1611.03184}{{\ttfamily 1611.03184}}].

\bibitem{Perez:2007emg}
F.~P\'erez and B.~E. Granger, \emph{{IPython: A System for Interactive
  Scientific Computing}},
  \href{https://doi.org/10.1109/MCSE.2007.53}{\emph{Comput. Sci. Eng.}
  {\bfseries 9} (2007) 21}.

\bibitem{Hunter:2007ouj}
J.~D. Hunter, \emph{{Matplotlib: A 2D Graphics Environment}},
  \href{https://doi.org/10.1109/MCSE.2007.55}{\emph{Comput. Sci. Eng.}
  {\bfseries 9} (2007) 90}.

\bibitem{SciPy}
E.~Jones, T.~Oliphant, P.~Peterson et~al., \emph{{SciPy}: Open source
  scientific tools for {Python}},  2001--.

\end{thebibliography}\endgroup

\end{document}